\documentclass[11pt]{article}
\pdfoutput=1
\usepackage{jheppub}
\usepackage{tabu}
\usepackage[vcentermath]{youngtab}
\usepackage[usenames,dvipsnames,table]{xcolor}
\usepackage{graphicx,amsmath,amssymb,amsthm,multirow,array,bm,bbm,esint}
\usepackage[mathscr]{eucal}
\usepackage[bbgreekl]{mathbbol}
\usepackage{epsf,amsfonts}
\usepackage{slashed}
\usepackage[numbers,sort&compress]{natbib}
\usepackage{minibox}
\usepackage{rotating}
\usepackage{pdflscape}
\usepackage{array,tikz-cd}
\usepackage{subcaption}
\usepackage{framed}
\usepackage{romannum}

\usepackage{slashed}

\def \sgn{\mbox{sgn\,}}

\def \tr{\mbox{tr\,}}
\def \str{\mbox{str\,}}

\def \ber{\mbox{Ber}\! }

\newcommand{\Jmatrix}{\mathcal{J} }
\newcommand{\jacobian}{\mathfrak{J} }

\newcommand{\st}{{\mbox{st}} }      
\newcommand{\parity}{{\pi} }      
\newcommand{\pst}{{\sharp} }      

\newcommand{\smD}{{ \mathfrak{D} }}

\newcommand{\mP}{{ \mathbb{P} }}
\newcommand{\mK}{{ \mathbb{K} }}
\newcommand{\mD}{{ \mathbb{D} }}
\newcommand{\mQ}{{ \mathbb{Q} }}

\newcommand{\mS}{{ \mathbb{S} }}

\newcommand{\mstar }{{ \circledast }}

\newcommand{\sD}{{  \mathtt{D} }}

\newcommand{\dP}{{ \cal P }}
\newcommand{\dK}{{ \cal K }}
\newcommand{\dD}{{ \cal D }}
\newcommand{\dQ}{{ \cal Q }}

\newcommand{\dS}{{ \cal S }}

\newcommand{\oscA}{{ \mathscr{A} }}
\newcommand{\oscB}{{ \mathscr{B} }}
\newcommand{\mB}{{ \mathbb{B} }}

\newcommand{\ie}{\emph{i.e.}, }
\newcommand{\eg}{\emph{e.g.}, }

\makeatletter
\newcommand{\doubletilde}[1]{{%
  \mathpalette\double@tilde{#1}%
}}
\newcommand{\double@tilde}[2]{%
  \sbox\z@{$\m@th#1\tilde{#2}$}%
  \ht\z@=.9\ht\z@
  \tilde{\box\z@}%
}
\makeatother

\title{Supersymmetric SYK Model\\
:Bi-local Collective Superfield/Supermatrix Formulation
}

\author{Junggi Yoon}
\affiliation{International Centre for Theoretical Sciences (ICTS-TIFR), \\
Shivakote, Hesaraghatta Hobli, Bengaluru 560089, India.}
\emailAdd{junggi.yoon@icts.res.in}

\abstract{
We discuss the bi-local collective theory for the $\mathcal{N}=1,2$ supersymmetric Sachdev-Ye-Kitaev~(SUSY SYK) models. We construct a bi-local superspace, and formulate the bi-local collective superfield theory of the one-dimensional SUSY vector model. The bi-local collective theory provides systematic analysis of the SUSY SYK models. We find that this bi-local collective theory naturally leads to supermatrix formulation in the bi-local superspace. This supermatrix formulation drastically simplifies the analysis of the SUSY SYK models. We also study $\mathcal{N}=1$ bi-local superconformal generators in the supermatrix formulation, and find the eigenvectors of teh superconformal Casimir. We diagonalize the quadratic action in large $N$ expansion.
}

\begin{document}
\maketitle


\section{Introduction}
\label{sec:introduction}

The Sachdev-Ye-Kitaev (SYK) model was proposed in~\cite{Sachdev:1992fk} and recently has been studied vigorously not only in the context of AdS/CFT~\cite{Sachdev:2010um,kitaevfirsttalk,KitaevTalks,Sachdev:2015efa} but also in the context of non-Fermi liquids~\cite{parcollet1999non,georges2000mean,georges2001quantum}. The SYK model is a quantum mechanical model of $N$ fermions with disordered interaction. In large $N$ diagrammatics, the dominance of ``melonic'' diagram make the model solvable at strong coupling limit~\cite{Sachdev:2010um,kitaevfirsttalk,KitaevTalks,Sachdev:2015efa,Polchinski:2016xgd,Jevicki:2016bwu,Maldacena:2016hyu,Jevicki:2016ito}. Also, this model features emergent reparametrization symmetry in the strict strong coupling limit after disorder average~\cite{kitaevfirsttalk,KitaevTalks,Polchinski:2016xgd,Jevicki:2016bwu,Maldacena:2016hyu,Jevicki:2016ito}. This reparametrization symmetry is broken spontaneously and explicitly at strong but finite coupling limit, which leads to Schwarzian effective action for Pseudo-Nambu-Goldstone modes~\cite{kitaevfirsttalk,KitaevTalks,Polchinski:2016xgd,Jevicki:2016bwu,Maldacena:2016hyu,Jevicki:2016ito}. Due to this mode, the SYK model is maximally chaotic, and the Lyapunov exponent of out-of-time-ordered correlator saturates chaos bound~\cite{KitaevTalks,Maldacena:2016hyu}. The same feature has been found in unitary quantum mechanical model of fermi tensors without disorder~\cite{Witten:2016iux,Gurau:2016lzk,Klebanov:2016xxf,Ferrari:2017ryl,Itoyama:2017emp,Itoyama:2017xid,Narayan:2017qtw,Mironov:2017aqv,Klebanov:2017nlk}. In tensor models, the ``melonic'' diagrams also dominate in large $N$, which leads to maximal chaos like the SYK model~\cite{Witten:2016iux,Gurau:2016lzk,Klebanov:2016xxf,Narayan:2017qtw}. This maximal chaos~\cite{Shenker:2013pqa,Roberts:2014isa,Maldacena:2015waa} indicates that both quantum mechanical models could be dual to gravity theory near horizon limit of extremal black hole, and the dual models have been proposed to be dilaton gravity~\cite{Maldacena:2016upp,Jensen:2016pah}, Liouville theories~\cite{Mandal:2017thl} and 3D gravity~\cite{Das:2017pif}. Because of these attractive features, the generalizations of the SYK and the tensor models have been studied in various context (\eg random matrix behavior~\cite{Cotler:2016fpe,Garcia-Garcia:2016mno,Krishnan:2016bvg,Li:2017hdt,Krishnan:2017ztz,Kanazawa:2017dpd}, flavor~\cite{Gross:2016kjj,Gurau:2017xhf}, lattice generalization in higher dimensions~\cite{Gu:2016oyy,Berkooz:2016cvq,Banerjee:2016ncu,Turiaci:2017zwd,Berkooz:2017efq,Jian:2017unn,Gu:2017ohj,Chaudhuri:2017vrv,Chen:2017dbb}, Schwarzian effective action~\cite{Forste:2017kwy,Stanford:2017thb,Belokurov:2017eit,Mertens:2017mtv} and supersymmetry~\cite{Fu:2016vas,Peng:2016mxj}, massive field instead of random coupling~\cite{Nishinaka:2016nxg,Peng:2017kro}, higher point function~\cite{Gross:2017hcz} and $1/N$ corrections~\cite{Bonzom:2017pqs,Dartois:2017xoe}.)

Most generalizations of the SYK model share the same feature: bi-local in time space. This bi-local structure is naturally appears in SYK model because the SYK model is essentially a large $N$ vector model. One of the systematic analysis of such large $N$ models was introduced as collective field theory in~\cite{Jevicki:1979mb}, which captures invariant physical degrees of freedom and provides the effective action thereof. The collective field theory has successfully analyzed the large $N$ models in the context of AdS/CFT~\cite{Jevicki:1980zq,Jevicki:1980zg,Jevicki:1991yi,Jevicki:1993qn,Avan:1995sp,deMelloKoch:1996mj,deMelloKoch:2002nq,Das:2003vw,Jevicki:2013kma,Jevicki:2015irq}. Especially, a bi-local collective field theory for three-dimensional $U(N)/O(N)$ vector model gave rich understanding of higher spin AdS$_4$/CFT $_3$ correspondence~\cite{Koch:2010cy,Jevicki:2011ss,Jevicki:2011aa,deMelloKoch:2012vc,Jevicki:2012fh,Das:2012dt,Jevicki:2014mfa,Koch:2014aqa,Koch:2014mxa,Jevicki:2015sla,Jevicki:2015pza}. However, in collective field theory, the bi-local structure is not restricted to space-time. In general, one can construct bi-local space of other abstract space in addition to spacetime. For example, in the bi-local thermofield CFT~\cite{Jevicki:2015sla}, the bi-local field is given by $\Psi(x_1,a;x_2,b)$ where $x_1$ and $x_2$ corresponds to spacetime as usual. $a,b(=1,2)$ represents labels of two copies of system in thermofield CFT, which corresponds to CFT lives on the left and right boundary of eternal black hole. Furthermore, we have also constructed bi-local field $\Psi(\tau_1,a;\tau_2,b)$ from the time ($\tau_1,\tau_2$) and replica space ($a,b=1,2,\cdots,n$) in the SYK model~\cite{Jevicki:2016bwu}.

In this paper, we will develop the bi-local collective superfield theory\footnote{Note that the collective theory for large $N$ supermatrix model was already studied in~\cite{Jevicki:1991yk,Rodrigues:1992by,deMelloKoch:1994ir}.} for one-dimensional vector model by constructing bi-local superspace, especially will focus on supersymmetric SYK model introduced by~\cite{Fu:2016vas}. This bi-local collective superfield theory enable us to analyze the effective action of SUSY SYK model in large $N$ systematically. Furthermore, in the bi-local collective theory, the matrix structure in the bi-local space naturally appears so that the bi-local collective theory can be seen as a matrix theory in the bi-local space. Hence, one can analyze the SUSY SYK model in the supermatrix formulation. This supermatrix formulation drastically simplifies analysis. We find that $\mathcal{N}=1$ superconformal generators becomes simple matrices in the supermatrix formulation. We also study the large $N$ classical solution and the large $N$ expansion of the collective action of the $\mathcal{N}=1$ SUSY SYK model. In particular, the quadratic action in large $N$ expansion can be easily diagonalized in the supermatrix formulations. Furthermore, the interaction term in the SUSY SYK model can be understood as the inner product in the supermatrix formulation. Furthermore, this also help diagonalize the rest of the quadratic action. We also emphasize that our formulation is not restricted to the SUSY SYK model. We develop a general framework to analyze large $N$ SUSY vector models as supermatrix theory in the bi-local superspace. Hence, this can be applied the generalization of the SUSY SYK models as well as other SUSY vector models.

The outline of the paper is as follows. In Section~\ref{sec: n=1 susy syk model}, we develop the bi-local collective superfield theory for one-dimensional $\mathcal{N}=1$ SUSY vector models, and we systematically study the collective superfield theory for $\mathcal{N}=1$ SUSY SYK model. $\mathcal{N}=1$ bi-local superconformal generators and eigenfunctions of superconformal Casimir is analyzed in Section~\ref{sec:bi-local superconformal symmetry}. In Section~\ref{sec:diagonalization of quadratic action}, using these eigenfunctions, we diagonalize the quadratic action of the collective action for $\mathcal{N}=1$ SUSY SYK model in large $N$. In Section~\ref{sec:n=2 syk model}, we also develop the bi-local collective superfield thoery for $\mathcal{N}=2$ SUSY vector models and discuss its application to SYK model. In Section~\ref{sec:conclusion}, we give our conclusion and future work.

\emph{Note added}: While this draft was under preparation, a related article~\cite{Murugan:2017eto,Peng:2017spg} appeared in arXiv.

\section{$\mathcal{N}=1$ Supersymmetric SYK Model }
\label{sec: n=1 susy syk model}

\subsection{Bi-local Superspace, Superfield and Supermatrix}
\label{sec: bi-local superspace and supermatrix}

Let us start with doubling the superspace $(\tau,\theta)$ to construct bi-local superspace:
\begin{equation}
(\tau,\theta)\quad \longrightarrow \quad (\tau_1,\theta_1;\tau_2,\theta_2)
\end{equation}
In this super bi-local space, superfields $A$ can be expanded as
\begin{equation}
A(\tau_1,\theta_1;\tau_2,\theta_2)\equiv A_0(\tau_1,\tau_2)+\theta_1 A_1(\tau_1,\tau_2)- A_2(\tau_1,\tau_2)\theta_2 -\theta_1 A_3(\tau_1,\tau_2)\theta_2
\end{equation}
where the lowest component $A_0$ could be either Grassmannian even or odd. This choice of the signs and the ordering of Grassmann variables will lead to a natural definition of a supermatrix and its multiplication. Furthermore, it is useful to call the superfield $A$ to be Grassmannian~odd (or, even) if the component $A_1$ and $A_2$ are Grassmannian~odd (or, even, respectively). \ie
\begin{equation}
A^\mp(\tau_1,\theta_1;\tau_2,\theta_2)=A^\pm_0(\tau_1,\tau_2)+\theta_1 A^\mp_1(\tau_1,\tau_2) - A^\mp_2(\tau_1,\tau_2)\theta_2 - \theta_1A^\pm_3(\tau_1,\tau_2) \theta_2
\end{equation}
Note that the lowest component of Grassmannian odd superfield is a Grassmannian even and vice versa. We will see later that this unusual definition is related to the fact that the star product (matrix multiplication) in the bi-local superspace is a Grassmannian odd operation.

Now, we define a star product (matrix multiplication) $\mstar$ in the bi-local superspace of two superfields $A$ and $B$ by
\begin{equation}
(A\mstar B)(\tau_1,\theta_1;\tau_2,\theta_2)\equiv\int  \; A(\tau_1,\theta_1;\tau_3,\theta_3)d\tau_3 d\theta_3 B(\tau_3,\theta_3;\tau_2,\theta_2)
\end{equation}
where the star product $\star$ of the components fields is the usual matrix multiplication of the bi-local space $(\tau_1,\tau_2)$. \ie $(A_i \star B_j)(\tau_1,\tau_2)\equiv \int d\tau_3\; A_i(\tau_1,\tau_3)B_j(\tau_3,\tau_2) $. Note that we place the (Grassmannian odd) measure between the two superfields to obtain a consistent star product $\mstar$ for all superfields. For example, the star product of two Grassmannian odd superfields is 
\begin{align}
&(A^-\mstar B^-)(\tau_1,\theta_1;\tau_2,\theta_2)\equiv\int  \; A^-(\tau_1,\theta_1;\tau_3,\theta_3)d\tau_3 d\theta_3 B^-(\tau_3,\theta_3;\tau_2,\theta_2)\cr
=&(A^+_0\star B^-_1 + A^-_2\star B^+_0) + \theta_1(A^-_1\star B^-_1+ A^+_3\star B^+_0) - (A^+_0\star B^+_3 + A^-_2\star B^-_2)\theta_2\cr
&- \theta_1( A^-_1\star B^+_3 + A^+_3\star B^-_2)\theta_2
\end{align}

This star product in bi-local superspace simplifies in the supermatrix formulation. We represent the superfields $A$ as a supermatrix as follow. \ie
\begin{equation}
A^\mp\equiv \begin{pmatrix}
A^\mp_1 & A^\pm_3 \\
A^\pm_0 & A^\mp_2
\end{pmatrix}
\end{equation}
In this definition of supermatrix, Grassmannian odd (even) superfield corresponds to Grassmannian odd (even) supermatrix. \eg
\begin{equation}
A=\underbrace{A_0}_{\text{Grassmannian Odd (Even)}} + \theta_1 \underbrace{A_1}_{\text{Grassmannian Even (Odd)}}+\cdots \;\Longleftrightarrow \; \underbrace{\begin{pmatrix}
A_1 & A_3\\
A_0 & A_2\\
\end{pmatrix}}_{\text{Grassmannian Even (Odd)}}
\end{equation}
Then, the star product $\mstar$ in the bi-local superspace becomes a simple matrix product:
\begin{equation}
(A\mstar B)(\tau_1,\theta_1;\tau_2,\theta_2)
=\begin{pmatrix}
A_1 & A_3 \\
A_0 & A_2
\end{pmatrix}\mstar \begin{pmatrix}
B_1 & B_3 \\
B_0 & B_2\\
\end{pmatrix}
\end{equation}
where the multiplication between component fields is the star product $\star$ in the bi-local space $(\tau_1,\tau_2)$. One can easily see that the identity supermatrix gives the expected delta function in the bi-local superspace. \ie
\begin{equation}
\mathbb{I}(\tau_1,\theta_1;\tau_2,\theta_2)\equiv \begin{pmatrix}
\delta(\tau_1-\tau_2) & 0 \\
0 & \delta(\tau_1-\tau_2)\\
\end{pmatrix}=(\theta_1-\theta_2) \delta(\tau_1-\tau_2)
\end{equation}
Furthermore, the natural definition of the trace in the bi-local superspace is consistent with the supertrace of a supermatrix. \ie
\begin{equation}
\int  d\tau_1 d\theta_1 \delta(\tau_{12}) \left[ A_0(\tau_1,\tau_2)+ \theta_1  A_1(\tau_1,\tau_2)-  A_2(\tau_1,\tau_2)\theta_1  \right]=  \tr A_1 - (-1)^{|A|}\tr A_2=\str A
\end{equation}
where $(-1)^{|A|}$ is $1$ if the supermatrix $A$ is Grassmannian even and $(-1)^{|A|}$ is $-1$ if $A$ is Grassmannian odd. Also, it is useful to define the superdeterminant (Berezinian) of the supermatrix. For our formulation, since the supermatrix is not restrict to be Grassmannian even, the supermatrix is defined by  
\begin{align}
&\ber (A) = \ber \begin{pmatrix}
A_1 & A_3 \\
A_0 &  A_2
\end{pmatrix}\cr
\equiv &\begin{cases}
\;\ber (A)=\det (A_1-A_3 A_2^{-1}A_0) \det(A_2)^{-1} & (A \mbox{: Grassmannian even})\\
\;\ber (\Jmatrix A)=\det (A_0-A_2 A_3^{-1}A_1) \det(-A_3)^{-1} &( A \mbox{: Grassmannian odd})\\
\end{cases}
\end{align}
where the constant supermatrix $\Jmatrix$ is defined by
\begin{equation}
\Jmatrix\equiv\begin{pmatrix}
0 & I\\
-I & 0\\
\end{pmatrix}\label{def: Jmatrix}\ .
\end{equation}

\subsection{Calculus of Bi-local Collective Superfield and Supermatrix}
\label{sec: n=2 calculus}

Before formulating the bi-local collective superfield theory, we clarify our conventions for the calculus of superfields. First of all, we define the functional derivatives of the same superfield by
\begin{equation}
{\delta f(\tau, \theta)\over \delta f(\tau',\theta')}\equiv (\theta'-\theta)\delta(\tau'-\tau)\label{def: fundamental field delta function}
\end{equation}
We also define a change of variables and a chain rule for a superfield:
\begin{align}
\delta f(\tau,\theta)\equiv& \int \delta g(\tau',\theta') d\tau' d\theta'  {\delta f(\tau, \theta)\over \delta g(\tau',\theta')}\\
{\delta\over \delta f(\tau,\theta)}\equiv & \int {\delta  g(\tau', \theta') \over \delta f(\tau,\theta)}d \tau' d\theta' {\delta\over \delta g(\tau', \theta')}
\end{align}
Note that we chose this unusual position of the Grassmannian odd measure to allow for uniform formulation independent of whether $f$ and $g$ are Grassmannian odd or even. This can easily be generalized to the bi-local collective superfields which could be Grassmannian odd or even. For example, one can check that this definition is consistent with the change of variables and the chain rule:
\begin{align}
\delta f_\alpha(\tau,\theta)=&\sum_\beta \int \delta f_\beta(\tau',\theta') d\tau' d\theta'  {\delta f_\alpha(\tau, \theta)\over \delta f_\beta(\tau',\theta')}=\int \delta f_\alpha(\tau',\theta') d\tau' d\theta'   (\theta'-\theta)\delta(\tau'-\tau)\\
{\delta\over \delta f_\alpha(\tau,\theta)}=&\sum_\beta\int {\delta  f_\beta(\tau', \theta') \over \delta f_\alpha(\tau,\theta)}d \tau' d\theta' {\delta\over \delta f_\beta(\tau', \theta')}=\int (\theta-\theta')\delta(\tau-\tau') d \tau' d\theta' {\delta\over \delta f_\alpha(\tau', \theta')}
\end{align}
where $\alpha$ runs over some complete basis. 

Furthermore, let us consider a change of variables and a chain rule for the bi-local superfield. In general, it is natural to define
\begin{equation}
{\delta F(\tau_1,\theta_1;\tau_2,\theta_2)\over \delta F(\tau_3,\theta_3;\tau_4,\theta_4)}\equiv (\theta_3-\theta_1)(\theta_4-\theta_2)\delta(\tau_3-\tau_1)\delta(\tau_4-\tau_2)
\end{equation}
Note that the RHS could be different depending on the symmetry of a superfield or supermatrix. Also, we find that the following convention for the change of variables and the chain rule of the bi-local superfield is consistent.
\begin{align}
\delta F(\tau_1,\theta_1;\tau_2,\theta_2)=& \int  \delta G(\tau_3,\theta_3;\tau_4,\theta_4) d\tau_4d\theta_4 d\tau_3 d\theta_3 {\delta F(\tau_1,\theta_1;\tau_2,\theta_2)\over \delta G(\tau_3,\theta_3;\tau_4,\theta_4)}\label{eq: change of variable}\\
{\delta \over \delta F(\tau_1,\theta_1;\tau_2,\theta_2)}= & \int {\delta G(\tau_3,\theta_3;\tau_4,\theta_4) \over \delta F(\tau_1,\theta_1;\tau_2,\theta_2)} d\tau_4 d\theta_4  d\tau_3 d\theta_3  {\delta \over \delta G(\tau_3,\theta_3;\tau_4,\theta_4) }\label{eq: chain rule}
\end{align}
For example, in this notation, we have
\begin{align}
&\delta((F\mstar G)(\tau_1,\theta_1;\tau_2,\theta_2))
= (\delta F\mstar G )(\tau_1,\theta_1;\tau_2,\theta_2)+ ( F\mstar \delta G )(\tau_1,\theta_1;\tau_2,\theta_2)
\end{align}
%
%
%

%

\subsection{Bi-local Collective Superfield Theory: Jacobian}
\label{sec:jacobian}

For the collective action for the SUSY vector model(\eg supersymmetric SYK models), we first study the Jacobian which appears in the transformation from the fundamental superfield to the bi-local collective superfield. Let us consider a superfield in $\mathcal{N}=1$ SUSY SYK model:
\begin{equation}
\psi^i(\tau,\theta)\equiv \chi^i(\tau)+\theta b^i(\tau)\hspace{1cm} (i=1,2,\cdots N)
\end{equation}
where $\chi^i$ is a Majorana fermion, and $b^i$ is a boson. This superfield transforms in the fundamental representation of $O(N)$:
\begin{equation}
\psi^i(\tau,\theta)  \quad \longrightarrow \quad O^{ij} \psi^j(\tau,\theta)
\end{equation}
It is natural to define a bi-local collective superfield which is invariant under $O(N)$ by 
\begin{equation}
\Psi(\tau_1,\theta_1;\tau_2,\theta_2)\equiv {1\over N} \psi^i(\tau_1,\theta_1) \psi^i(\tau_2,\theta_2)
\end{equation}
It is important to note that the bi-local superfield is anti-symmetric in the bi-local superspace. \ie
\begin{equation}
\Psi(\tau_1,\theta_1;\tau_2,\theta_2)=-\Psi(\tau_2,\theta_2;\tau_1,\theta_1)\label{eq:anti-sym of bilocal}
\end{equation}
When changing variables in the path integral from the fundamental superfield to bi-local collective superfield, we will get a non-trivial Jacobian. To obtain the Jacobian, it is useful to consider the following identity for an arbitrary functional $F[\Psi]$.
\begin{equation}
\sum_{i}\int \mathcal{D}\psi {\delta\over \delta \psi^i(\tau_1,\theta_1)}\left[ \psi^i(\tau_2,\theta_2) F[\Psi] e^{-S}\right]=0\label{eq:functional identity1}
\end{equation}
Using the chain rule of the bi-local superfield in~\eqref{eq: chain rule}, we have
\begin{align}
&\psi^i(\tau_2,\theta_2){\delta\over \delta \psi^i(\tau_1,\theta_1)}= \int \psi^i(\tau_2,\theta_2) {\delta \Psi(\tau_3,\theta_3;\tau_4,\theta_4)\over \delta \psi^i(\tau_1,\theta_1)}d\tau_4 d\theta_4 d\tau_3 d\theta_3 {\delta \over \delta  \Psi(\tau_3,\theta_3;\tau_4,\theta_4)}\cr
=&2\int \Psi(\tau_2,\theta_2;\tau_3,\theta_3)d\tau_3 d\theta_3 {\delta \over \delta  \Psi(\tau_1,\theta_1;\tau_3,\theta_3)}
\end{align}
Hence, recalling our convention~\eqref{def: fundamental field delta function}, \eqref{eq:functional identity1} can be written as
\begin{align}
&N (\theta_1-\theta_2)\delta(\tau_1-\tau_2)\langle F\rangle +2\int  \left\langle\Psi(\tau_2,\theta_2;\tau_3,\theta_3) d \tau_3 d\theta_3 {\delta F[\Psi]\over \delta \Psi(\tau_1,\theta_1;\tau_3,\theta_3)}\right\rangle\cr
&-2\int  \left\langle  \Psi(\tau_2,\theta_2;\tau_3,\theta_3)d \tau_3 d\theta_3 {\delta S[\Psi]\over \delta \Psi(\tau_1,\theta_1;\tau_3,\theta_3)}F[\Psi] \right\rangle=0\label{eq: identity for jacobian 1}
\end{align}
where we used the fact that the superfield ${\delta\over \delta \psi^i(\tau,\theta)}$ is Grassmannian even. 

On the other hand, one can also utilize a similar identity in the bi-local collective representation:
\begin{equation}
\int \mathcal{D}\psi \int d\tau_3 d\theta_3 {\delta\over \delta \Psi(\tau_1,\theta_1;\tau_3,\theta_3)}\left[ \Psi(\tau_2,\theta_2;\tau_3,\theta_3)\; \jacobian\; F[\Psi] e^{-S}\right]=0\label{eq:functional identity2}
\end{equation}
where $\jacobian=\jacobian[\Psi]$ is the Jacobian for the bi-local collective representation. Then, we have 
\begin{align}
&{1\over 2}(\theta_1-\theta_2)\delta(\tau_1-\tau_2)\langle F[\Psi]\rangle +\int \left\langle \Psi(\tau_2,\theta_2;\tau_3,\theta_3) d\tau_3 d\theta_3 {\delta \log \jacobian \over \delta \Psi (\tau_1,\theta_1;\tau_3,\theta_3) }  \; F[\Psi] \right\rangle\cr
&+\int \left\langle \Psi(\tau_2,\theta_2;\tau_3,\theta_3) d\tau_3 d\theta_3   {\delta F \over \delta \Psi (\tau_1,\theta_1;\tau_3,\theta_3) } \right\rangle\cr
&-\int  \left\langle \Psi(\tau_2,\theta_2;\tau_3,\theta_3)d\tau_3 d\theta_3  {\delta S\over \delta \Psi (\tau_1,\theta_1;\tau_3,\theta_3) }  \; F[\Psi] \right\rangle=0\label{eq: identity for jacobian 2}
\end{align}
Note that we used
\begin{align}
{\delta \Psi(\tau_1,\theta_1;\tau_2,\theta_2)\over \delta \Psi(\tau_3,\theta_3;\tau_4,\theta_4)}\equiv &{1\over 2}(\theta_3-\theta_1)(\theta_4-\theta_2)\delta(\tau_3-\tau_1)\delta(\tau_4-\tau_2)\cr
&-{1\over 2}(\theta_3-\theta_2)(\theta_4-\theta_1)\delta(\tau_3-\tau_2)\delta(\tau_4-\tau_1)\label{eq:functional derivative of bi-local}
\end{align}
which is imposed by anti-symmetry of the bi-local superfield~$\Psi$ in~\eqref{eq:anti-sym of bilocal}. As usual in supersymmetry, we do not have divergence proportional to $\delta(\tau-\tau)$ unlike what appears in the bosonic bi-local collective field theory~\cite{deMelloKoch:1996mj,Das:2003vw,Koch:2010cy,Jevicki:2014mfa}. In our formulation, this naturally comes from the fact that the analogous $(\theta-\theta)\delta(\tau-\tau)$ for superspace, vanishes.  From \eqref{eq: identity for jacobian 1} and \eqref{eq: identity for jacobian 2} for an arbitrary functional of $F[\Psi]$, we obtain a functional differential equation for the Jacobian $\jacobian$:
\begin{equation}
{N-1\over 2} (\theta_1-\theta_2) \delta(\tau_1-\tau_2)=\int  \Psi(\tau_2,\theta_2;\tau_3,\theta_3) d\tau_3 d\theta_3 {\delta \log \jacobian \over \delta \Psi (\tau_1,\theta_1;\tau_3,\theta_3) } \label{eq:SD equation for Jacobian}
\end{equation}
This differential equation can easily be solved using the supermatrix formulation in Section~\ref{sec: bi-local superspace and supermatrix}. In the supermatrix formulation, it is trivial to conclude that
\begin{equation}
\log \jacobian=-{N-1\over 2} \str \log \Psi(\tau_1,\theta_1;\tau_2,\theta_2)\label{eq:jacobian result}
\end{equation}
We emphasize that anti-symmetry of the bi-local superfield\footnote{We thank to Robert de Mello Koch for pointing out this.} leads to a term ${1\over 2}(\theta_1-\theta_2)\delta(\tau_1-\tau_2) $ in \eqref{eq: identity for jacobian 2}, which shifts large $N$ to $N-1$. This shift of large $N$ in the Jacobian was already observed in non-supersymmetric bi-local collective field theory~\cite{Jevicki:2014mfa}, and it was shown to play an important role in matching one-loop free energies of higher spin theories and vector models~\cite{Giombi:2013fka,Jevicki:2014mfa,Giombi:2014iua,Giombi:2014yra,Giombi:2016pvg}. Though this shift is not crucial for the discussion in this paper, it is essential to obtain the exact result. For example, one can consider a free one-dimensional $\mathcal{N}=1$ SUSY vector model for which one knows the exact answer.\footnote{We also thank to Robert de Mello Koch for raising this issue and confirming the result.} We confirm that the shift $N-1$ gives the correct one-point function of bi-local superfield (or, invariant two-point function of fundamental superfields) (See Appendix~\ref{app: correction}).

\subsection{Bi-local Collective Superfield Theory for $\mathcal{N}=1$ SUSY SYK Model}
\label{}

In~\cite{Fu:2016vas}, the action of the supersymmetric SYK model is given by
\begin{equation}
\mathcal{L}=\sum_i\left[{1\over 2} \chi^i\partial \chi^i-{1\over 2} b^i b^i + i\sum_{1\leqq j < k \leqq N} C_{ijk}b^i \chi^j \chi^k\right]
\end{equation}
%
%
%
where $C_{ijk}$ is a random coupling constant, and is totally anti-symmetric in its indices. After the disorder average of the random coupling constant $C_{ijk}$ over a Gaussian distribution\footnote{Rigorously, we perform annealed average instead of a quenched average. For a proper quenched average, one has to use the replica trick, which was also done for non-supersymmetric bi-local collective field theory in~\cite{Jevicki:2016bwu}.}, one has an effective action~\cite{Fu:2016vas}:
\begin{align}
S_{\text{eff}}=&\int  d\tau \left( {1\over 2} \chi^i\partial \chi^i-{1\over 2} b^i b^i \right)-{J\over 2 N^2}\int  d\tau_1 d\tau_2 [b^i(\tau_1) b^i(\tau_2)][\chi^j(\tau_1) \chi^j(\tau_2)]^2\cr
&-{J\over N^2}\int  [b^i(\tau_1)\chi^i(\tau_2)][\chi^j(\tau_1)b^j(\tau_2)][\chi^k(\tau_1)\chi^k(\tau_2)]\ .
\end{align}
Note that the disorder average leads to an emergent $O(N)$ symmetry. As before, we define the (fundamental) superfield by
\begin{equation}
\psi^i(\tau,\theta)\equiv \chi^i(\tau)+\theta b^i(\tau)
\end{equation}
we will express the effective action in terms of the bi-local collective superfield given by
\begin{align}
&\Psi(\tau_1,\theta_1;\tau_2,\theta_2)\equiv {1\over N} \sum_{i=1}^N \psi^i(\tau_1,\theta_1) \psi^i(\tau_2,\theta_2)\cr
=&{1\over N}\sum_{i=1}^N \left[\chi^i(\tau_1)\chi^i(\tau_2)+\theta_1 b^i(\tau_1)\chi^i(\tau_2) + \chi^i(\tau_1) b^i(\tau_2) \theta_2 + \theta_1 b^i(\tau_1) b^i(\tau_2) \theta_2\right]
\end{align}
In terms of supermatrix notation, the bi-local superfield can be represented as
\begin{equation}
\Psi(\tau_1,\theta_1;\tau_2,\theta_2)={1\over N} \sum_{i=1}^N\begin{pmatrix}
b^i(\tau_1)\chi^i(\tau_2)  & -b^i(\tau_1) b^i(\tau_2) \\
\chi^i(\tau_1)\chi^i(\tau_2) & - \chi^i(\tau_1) b^i(\tau_2)\\
\end{pmatrix}
\end{equation}
Recall that the bi-local superfield is anti-symmetric in the bi-local superspace (See \eqref{eq:anti-sym of bilocal}.) As a supermatrix, the bi-local supermatrix has the following symmetry. \ie
\begin{equation}
\Jmatrix\Psi^\st\Jmatrix= \Psi \label{eq: antisymmetry of supermatrix}
\end{equation}
where $A^\st$ is the supertranspose of a supermatrix $A$ defined by
\begin{equation}
A^\st\equiv \begin{pmatrix}
A_1^t & (-1)^{|A|}A_0^t\\
-(-1)^{|A|}A_3^t & A_2^t\\
\end{pmatrix}
\end{equation}
and the matrix $\Jmatrix$ is given in \eqref{def: Jmatrix}.

For the collective action, it is useful to define a superderivative matrix:
\begin{align}
\smD(\tau_1,\theta_1;\tau_2,\theta_2)\equiv& \sD_{\theta_1}(\theta_1-\theta_2)\delta(\tau_1-\tau_2)= \delta(\tau_1-\tau_2) -\theta_1\partial_{\tau_1}\delta(\tau_1-\tau_2) \theta_2\cr
 =& \begin{pmatrix}
0 & \partial_1\delta(\tau_1-\tau_2) \\
\delta(\tau_1-\tau_2) & 0\\
\end{pmatrix}\label{def:bi-local superderivative}
\end{align}
where the superderivative $\sD_{\theta_1}$ is defined by
\begin{align}
\sD_{\theta_1}\equiv&\partial_{\theta_1} + \theta_1\partial_{\tau_1} 
\end{align}
Note that the superderivative matrix $\smD$ is Grassmannian odd supermatrix. Using the supermatrix formulation, one can easily check that
\begin{align}
( \smD\mstar A)(\tau_1,\theta_1;\tau_2,\theta_2)=\begin{pmatrix}
\partial_{\tau_1}A_0(\tau_1,\tau_2) & \partial_{\tau_1}A_2(\tau_1,\tau_2)\\
A_1 & A_3\\
\end{pmatrix}
\end{align}
and, therefore, the supertrace of the supermatrix leads to the kinetic term:
\begin{equation}
\str ( \mathbb{D}\mstar \Psi)=\int d\tau_1 \left[\left.\partial_{\tau_1}\psi^i (\tau_1)\psi^i(\tau_2)\right|_{\tau_2\rightarrow \tau_1 } + b^i(\tau_1) b^i(\tau_1)\right]
\end{equation}
As an aside, the superderivative matrix has a similar property as the ordinary superderivative. \ie
\begin{equation}
(\smD\mstar\smD)(\tau_1,\theta_1;\tau_2,\theta_2)=\partial_{\tau_1}\begin{pmatrix}
\delta(\tau_1-\tau_2) & 0\\
0 & \delta(\tau_1-\tau_2) \\
\end{pmatrix}= \partial_{\tau_1} \mathbb{I}(\tau_1,\theta_1;\tau_2,\theta_2)
\end{equation}
where $\mathbb{I}(\tau_1,\theta_1;\tau_2,\theta_2)$ is the identity supermatrix. Hence, one can immediately obtain the bi-local collective action for the SUSY SYK model.
\begin{equation}
S_{col}=-{N\over 2} \str \left[\smD\mstar  \Psi\right]+{N\over 2}\str \log \Psi - {JN\over 6}\int  d\tau_1d\theta_1 d\tau_2 d\theta_2 [\Psi(\tau_1,\theta_1;\tau_2,\theta_2)]^3\label{eq: bi-local collective action for SYK model2}
\end{equation}
Also, one can rewrite the collective action completely in terms of supermatrix notation.
\begin{equation}
S_{col}={N\over 2} \str \left[ - \smD\mstar  \Psi+ \log \Psi - {J\over 3} \Psi  \mstar [\Psi]^2 \right]\label{eq: bi-local collective action for SYK model}
\end{equation}
where we define $[\Psi]^2(\tau_1,\theta_1;\tau_2,\theta_2)\equiv [\Psi(\tau_1,\theta_1;\tau_2,\theta_2)]^2 $. Note that it is also straightforward to generalize this into general $q$ case, which we present in Section~\ref{app: general q}. Note that in this paper we drop the shift in $N$ found in~\eqref{eq:jacobian result} for simplicity because it does have an effect on our discussions. But, one should take this into account for the sub-leading calculations in $1/N$.

\subsection{Large $N$ Classical Solution}
\label{}

At large $N$, the variation with respect to the bi-local superfield gives the large $N$ classical solution. Note that in the supermatrix notation, the variation of the collective action~\eqref{eq: bi-local collective action for SYK model} can easily be performed.\footnote{It is sometimes simpler to vary the collective action in terms of superfield notation. For instance, the variation of the third term in \eqref{eq: bi-local collective action for SYK model2} can be expressed as
\begin{align}
{JN\over 2}\int  d\tau_1d\theta_1 d\tau_2 d\theta_2 \;\delta\Psi(\tau_1,\theta_1;\tau_2,\theta_2) [\Psi(\tau_1,\theta_1;\tau_2,\theta_2)]^2={JN\over 2} \str\left( \delta \Psi\mstar [\Psi]^2 \right)
\end{align}
}
Hence, one can immediately obtain the large $N$ saddle-point equation of the collective action:
\begin{equation}
- \smD +  \Psi^{-1} - J \Psi^2=0
\end{equation}
or equivalently, by multiplying supermatrix $\Psi$, we have
\begin{equation}
-  \smD\mstar \Psi +  \mathbb{I} - J [\Psi^2]\mstar \Psi=0\label{eq: large n classical equation}
\end{equation}
The most general ansatz for a scaling solution is given~\cite{Fu:2016vas} by
\begin{align}
&\Psi_{cl}(\tau_1,\theta_1;\tau_2,\theta_2)={c_1\sgn(\tau_{12}-\theta_1\theta_2) \over |\tau_{12}-\theta_1\theta_2|^{2\Delta_1}} +\theta_{12}{c_2+c_3\sgn(\tau_{12})\over |\tau_{12}-\theta_1\theta_2|^{2\Delta_2} }\cr
=&{c_1\over |\tau_{12}|^{2\Delta_1}} \left[ \sgn(\tau_{12})+ 2\Delta_1{\theta_1\theta_2\over  |\tau_{12}| }  \right]+ \theta_{12}{c_2+c_3\sgn(\tau_{12})\over |\tau_{12}|^{2\Delta_2}}=\begin{pmatrix}
{c_2+c_3\sgn(\tau_{12})\over |\tau_{12}|^{2 \Delta_2} } & - {2 \Delta_1c_1\over |\tau_{12}|^{2 \Delta_1+1} } \\
{c_1\sgn(\tau_{12})\over |\tau_{12}|^{2 \Delta_1} } &- {c_2+c_3\sgn(\tau_{12})\over |\tau_{12}|^{2 \Delta_2} } \\
\end{pmatrix}\cr
\equiv& \begin{pmatrix}
c_2 f^s_{2\Delta_2}(\tau_{12})+c_3 f^a_{2\Delta_2}(\tau_{12}) & -2 \Delta_1 c_1 f^s_{2\Delta_1+1}(\tau_{12})\\
c_1 f^a_{2\Delta_1}(\tau_{12}) & -c_2 f^s_{2\Delta_2}(\tau_{12})-c_3 f^a_{2\Delta_2}(\tau_{12}) \\
\end{pmatrix}
\end{align}
where we define $\theta_{12}=\theta_1-\theta_2$ and
\begin{equation}
f^s_\mu(\tau)\equiv {1\over |\tau|^\mu}\;\;,\qquad f^a_\mu(\tau)\equiv {\sgn(\tau)\over |\tau|^\mu}
\end{equation}
Note that $c_1$ is Grassmannian even while $c_2$ and $c_3$ are Grassmannian odd. Moreover, $[\Psi]^2$ can also be expressed as a supermatrix:
\begin{align}
&[\Psi_{cl}]^2(\tau_1,\theta_1;\tau_2,\theta_2)={c_1^2\over |\tau_{12}|^{4\Delta_1}} \left[ 1+  \theta_1\theta_2{4\Delta_1\sgn(\tau_{12})\over |\tau_{12}| }  \right]+\theta_{12}{2c_1(c_2\sgn(\tau_{12}) +c_3)\over |\tau_{12}|^{2\Delta_1+2\Delta_2}}\cr
=&\begin{pmatrix}
2c_1\left[c_2 f^a_{ 2\Delta_1+2\Delta_2}(\tau_{12}) +c_3 f^s_{2\Delta_1+2\Delta_2}(\tau_{12}) \right]   & -4\Delta_1 c_1^2 f^a_{4\Delta_1+1}(\tau_{12})  \\
c_1^2 f^s_{4\Delta_1}(\tau_{12})  & -2c_1\left[c_2 f^a_{2\Delta_1+2\Delta_2}(\tau_{12}) + c_3 f^s_{2\Delta_1+2\Delta_2}(\tau_{12}) \right]  \\
\end{pmatrix}
\end{align}
%
%
Using the integrals
\begin{align}
\int  d\tau \; {1 \over |\tau|^\lambda}e^{i w\tau }=&2w^{\lambda-1} \int_0^\infty  dx \; x^{-\lambda} \cos x =2w^{\lambda-1} \Gamma(1-\lambda)\sin {\pi\lambda\over 2}\\
\int  d\tau \; {\sgn(\tau)  \over |\tau|^\lambda} e^{i w\tau } =& 2i w^{\lambda-1} \int_0^\infty dx \; x^{-\lambda} \sin x=2i w^{\lambda-1}\Gamma(1-\lambda) \cos {\pi \lambda\over 2}
\end{align}
we can Fourier transform $f^s_\lambda(\tau)$ and $f^a_\lambda(\tau)$ into $\tilde{f}^s_\lambda w^{\lambda-1}$ and $\tilde{f}^a_\lambda(w) w^{\lambda-1}$, respectively. In addition, one can write the star product of $f$'s in terms of $\tilde{f}^s_\lambda(w)$ and $\tilde{f}^a_\lambda(w)$ as follows
\begin{equation}
(f_\lambda^{p_1} \star f_\nu^{p_2} )(\tau_1,\tau_2) ={1\over 2\pi  } \int dw \; e^{-iw\tau_{12}} \tilde{f}_\lambda^{p_1} \tilde{f}_\nu^{p_2}  w^{\lambda+\nu-2}\hspace{1cm} (p_1,p_2=s,a)\label{eq: integral classical solution1}
\end{equation}
%
%
%
%
%
where
\begin{align}
\tilde{f}^s_\lambda\equiv 2\Gamma(1-\lambda)\sin{\pi \lambda\over 2}\;\;,\qquad \tilde{f}^a_\lambda\equiv 2i \Gamma(1-\lambda)\cos {\pi \lambda\over 2}\label{eq: integral classical solution2}
\end{align}
Thus, the third term in \eqref{eq: large n classical equation} can be written as
\begin{align}
 &[\Psi]^2\mstar \Psi(\tau_1,\theta_1;\tau_2,\theta_2)\cr
 =&{1\over 2\pi i } \int  dw\; e^{-iw\tau_{12}} \begin{pmatrix}
2c_1\left[c_2 \tilde{f}^a_{2\Delta_1+2\Delta_2}+c_3 \tilde{f}^s_{2\Delta_1+2\Delta_2} \right]   & -4\Delta_1 c_1^2 \tilde{f}^a_{4\Delta_1+1} \\
c_1^2 \tilde{f}^s_{4\Delta_1}  & -2c_1\left[c_2 \tilde{f}^a_{2\Delta_1+2\Delta_2} + c_3 \tilde{f}^s_{2\Delta_1+2\Delta_2} \right]  \\
\end{pmatrix}\cr
&\times
\begin{pmatrix}
c_2 \tilde{f}^s_{2\Delta_2}+c_3 \tilde{f}^a_{2\Delta_2} & -2 \Delta_1 c_1 \tilde{f}^s_{2\Delta_1+1}\\
c_1 \tilde{f}^a_{2\Delta_1} & -c_2 \tilde{f}^s_{2\Delta_2}-c_3 \tilde{f}^a_{2\Delta_2} \\
\end{pmatrix}\label{eq: classical equation potential}
 \end{align}
where the matrix multiplication in the integrand is ordinary matrix multiplication. Recalling the action of the bi-local superderivative, the first term of \eqref{eq: large n classical equation} becomes
\begin{align}
&\smD\mstar \Psi_{cl}=\begin{pmatrix}
\partial_{\tau_1} \Psi_{cl,0}(\tau_1,\tau_2) & \partial_{\tau_1} \Psi_{cl,2}(\tau_1,\tau_2)\\
\Psi_{cl,1}(\tau_1,\tau_2) & \Psi_{cl,3}(\tau_1,\tau_2)\\
\end{pmatrix}\cr
=&{1\over 2\pi i} \int dw \; e^{-iw\tau_{12}} \begin{pmatrix}
-iwc_1 \tilde{f}^a_{2\Delta_1} & iw[c_2 \tilde{f}^s_{2\Delta_2}+c_3 \tilde{f}^a_{2\Delta_2}] \\
c_2 \tilde{f}^s_{2\Delta_2}+c_3 \tilde{f}^a_{2\Delta_2} & -2 \Delta_1 c_1 \tilde{f}^s_{2\Delta_1+1}\\
\end{pmatrix}\label{eq: classical equation kinetic}
\end{align}
while the second term of \eqref{eq: large n classical equation} is trivially given by
\begin{equation}
\mathbb{I}={1\over 2\pi i} \int dw \; e^{-iw\tau_{12} } \begin{pmatrix}
1 & 0 \\
0 & 1 \\
\end{pmatrix}\label{eq: classical equation jacobian}
\end{equation}
Now, we will consider the strong coupling limit:
\begin{equation}
{w\over J} \ll 1 
\end{equation}
Note that the constants $c_1,c_2$ and $c_3$ should be scaled with $J$ as follows
\begin{equation}
c_1\sim J^{-2\Delta_1} \quad,\quad c_2\;,\; c_3\sim J^{-2\Delta_2+{1\over 2} }
\end{equation}
Requiring positive conformal dimensions, matching the power-laws of the diagonal elements of the classical equation~\eqref{eq: large n classical equation} gives 
\begin{equation}
\Delta_1={1\over 6} \qquad\mbox{or}\qquad 2\Delta_1+4\Delta_2=2
\end{equation}

Let us consider the first case. \ie
\begin{equation}
\Delta_1={1\over 6}
\end{equation}
We match the leading terms of the diagonal elements in the classical equation~\eqref{eq: large n classical equation}. In this case, the off-diagonal elements from $[\Psi]^2\mstar \Psi$ diverge in the strong coupling limit for $\Delta_2<{2\over 3}$. This divergence cannot be eliminated by tuning the coefficients. Moreover, for $\Delta_2>{2\over 3}$, these terms vanish in the strong coupling limit. However, since we want reparametrization symmetry in the strict strong coupling limit, we had better not treat $[\Psi]^2\mstar \Psi$ as a perturbation. Hence, we find that the only solution is given by
\begin{equation}
1-2\sqrt{3}\pi c_1^3J=0 \quad, \quad c_2=c_3=0
\end{equation}
Note that we do not have to find $\Delta_2$ because $c_2=c_3=0$. Also, note that the kinetic term $\smD\mstar \Psi$ is a perturbation in the strong coupling limit as in the non-supersymmetric SYK model.

Next, we analyze the second case. \ie
\begin{equation}
2\Delta_1+4\Delta_2=2
\end{equation}
For this case, the off-diagonal elements contain divergent terms of order $\mathcal{O}(w^{-\Delta_1})$ in the strong coupling limit. To remove this divergence, we choose
\begin{equation}
c_3=i c_ 2 \cot{\pi \Delta_1\over 2}
\end{equation}
But, in this case, one cannot solve the diagonal and off-diagonal classical solution simultaneously.

To summarize, the classical solution is found to be
\begin{equation}
\Psi_{cl}=c{\sgn(\tau_{12}-\theta_1\theta_2) \over |\tau_{12}-\theta_1\theta_2|^{1/3} }=c\begin{pmatrix}
0 & - {1\over 3 |\tau_{12}|^{4/3}} \\
{\sgn(\tau_{12})\over |\tau_{12}|^{1/3}} & 0\\
\end{pmatrix}\label{eq:classical solution q=3}
\end{equation}
where
\begin{equation}
1-2\sqrt{3}\pi c^3J=0
\end{equation}
This classical solution was already found in~\cite{Fu:2016vas}, and corresponds to a vacuum with definite fermion number.

\subsection{Large $N$ Expansion and Quadratic Action}
\label{sec: quadratic action N=1}

Now, we expand the collective action~\eqref{eq: bi-local collective action for SYK model} for the bi-local superfield:
\begin{equation}
\Psi(\tau_1,\theta_1;\tau_2,\theta_2)\equiv \Psi_{cl}(\tau_1,\theta_1;\tau_2,\theta_2)+\sqrt{2\over N} \Phi(\tau_1,\theta_1;\tau_2,\theta_2)
\end{equation}
where $\Phi(\tau_1,\theta_1;\tau_2,\theta_2)$ is a bi-local fluctuation around the classical solution $\Psi_{cl}$ given by
\begin{align}
\Phi(\tau_1,\theta_1;\tau_2,\theta_2)=&\varphi(\tau_1,\tau_2)+\theta_1\eta_1(\tau_1,\tau_2) -\eta_2(\tau_1,\tau_2)\theta_2 - \theta_1 \sigma(\tau_1,\tau_2)\theta_2\cr
=&\begin{pmatrix}
\eta_1(\tau_1,\tau_2) & \sigma(\tau_1,\tau_2)\\
\varphi(\tau_1,\tau_2) & \eta_2(\tau_1,\tau_2)\\
\end{pmatrix}
\end{align}
Note that the anti-symmetry of the bi-local field in~\eqref{eq:anti-sym of bilocal} leads to 
\begin{align}
\varphi(\tau_1,\tau_2)=&-\varphi(\tau_2,\tau_1)\\
\sigma(\tau_1,\tau_2)=&\sigma(\tau_2,\tau_1)\\
\eta_1(\tau_1,\tau_2)=&\eta_2(\tau_2,\tau_1)
\end{align}
or, equivalently, we have 
\begin{equation}
\Jmatrix \Phi^\st \Jmatrix= \Phi
\end{equation} 
From the supermatrix notation, one can easily obtain the quadratic action:
\begin{equation}
S_{col}^{(2)}=-{1\over 2} \str ( \Psi_{cl}^{-1}\mstar \Phi\mstar \Psi_{cl}^{-1} \mstar \Phi) - J \int d\tau_1d\theta_1d\tau_2 d\theta_2 \; \Psi_{cl}(\tau_1,\theta_1;\tau_2,\theta_2) [\Phi(\tau_1,\theta_1;\tau_2,\theta_2)]^2
\end{equation}
From the classical equation, the inverse supermatrix is given by
\begin{align}
\Psi_{cl}^{-1}(\tau_3,\theta_3;\tau_2,\theta_2)=&-J [\Psi_{cl}(\tau_3,\theta_3;\tau_2,\theta_2)]^2 
= Jc^2 \begin{pmatrix}
0 & -{2\over 3} f^a_{5/3}(\tau_{12}) \\
f^s_{2/3}(\tau_{12})  & 0 \\
\end{pmatrix}
\end{align}
Hence, one can write the kinetic term as 
\begin{align}
&{1\over 2} \str( \Psi_{cl}^{-1}\mstar \Phi \mstar \Psi_{cl}^{-1} \mstar \Phi)\cr
=&-{J^2c^4\over 2} \tr\left( {4\over 9}  f^a_{5/3}\star \varphi \star f^a_{5/3}\star \varphi -  f^s_{2/3}\star \sigma \star f^s_{2/3}\star \sigma + {4\over 3}  f^s_{2/3}\star \eta_1 \star f^a_{5/3}\star \eta_2 \right)
\end{align}
where the cross terms are cancelled because of the supertrace. Also, the classical solution can be written as
\begin{align}
\Psi_{cl}(\tau_1,\theta_1;\tau_2,\theta_2)=&c\left[{ \sgn(\tau_{12})\over\tau_{12}^{1/3}} +{\theta_1\theta_2\over 3\tau_{12}^{4/3} }\right]\equiv c\left[ f^a_{1/3}(\tau_{12})-\theta_1 \left(-{1\over 3}  f^s_{4/3}(\tau_{12})\right)\theta_2\right]\cr
=&c\begin{pmatrix}
0 & -{1\over 3}  f^s_{4/3}(\tau_{12}) \\
f^a_{1/3}(\tau_{12}) & 0  \\
\end{pmatrix}
\end{align}
The square of bi-local fluctuation can be also written using the supermatrix notation:
\begin{equation}
[\Phi(\tau_1,\theta_1;\tau_2,\theta_2)]^2
=\begin{pmatrix}
2[\varphi\eta_1](\tau_1,\tau_2) & 2[\varphi\sigma](\tau_1,\tau_2) +2[\eta_1\eta_2](\tau_1,\tau_2) \\
[\varphi^2](\tau_1,\tau_2) & 2[\varphi\eta_2](\tau_1,\tau_2) \\
\end{pmatrix}
\end{equation}
which leads to 
\begin{align}
&J\int d\tau_1 d\tau_2 d\theta_1 d\theta_2 \overline{ \Psi}(\tau_1,\theta_1;\tau_2,\theta_2) [\Phi]^2(\tau_1,\theta_1;\tau_2,\theta_2)\cr
=&Jc   \; \tr \left(  -{1\over 3} f^s_{4/3} \star[\varphi^2]- 2f^a_{1/3}\star [\varphi\sigma]-2f^a_{1/3}\star [\eta_1\eta_2] \right)
\end{align}
In conclusion, the quadratic action can be manipulated as follows.
\begin{align}
S^{(2)}=&{Jc\over 4\sqrt{3}\pi }\tr\left(- {4\over 9}  f^a_{5/3}\star \varphi \star f^a_{5/3}\star \varphi +  f^s_{2/3}\star \sigma \star f^s_{2/3}\star \sigma - {4\over 3}  f^s_{2/3}\star \eta_1 \star f^a_{5/3}\star \eta_2\right.\cr
&\hspace{3cm}\left. +{4\sqrt{3}\pi \over 3} f^s_{4/3} \star[\varphi^2] + 8\sqrt{3}\pi f^a_{1/3}\star [\varphi\sigma] + 8\sqrt{3}\pi f^a_{1/3}\star [\eta_1\eta_2] \right)\label{eq:quadratic action in component fields}
\end{align}
In the section~\ref{sec:diagonalization of quadratic action}, we will diagonalize this quadratic action. Though we express the quadratic action in terms of component fields for pedagogical purposes, we will not use this expression~\eqref{eq:quadratic action in component fields} in terms of component fields for the diagonalization of the quadratic action. Instead, we find that the collective action of $\mathcal{N}=1$ SUSY SYK model can completely be written in term of the supermatrix notation:
\begin{equation}
S_{col}^{(2)}=-{1\over 2} \str \left(  \Psi_{cl}^{-1}\mstar \Phi\mstar \Psi_{cl}^{-1} \mstar \Phi + 2J\Phi\mstar [\Psi_{cl}\Phi]\right)\label{eq:quadratic action supermatrix}
\end{equation}
We will see that It is much easier to diagonalize the quadratic action.

\section{$\mathcal{N}=1$ Bi-local Superconformal Algebra}
\label{sec:bi-local superconformal symmetry}

\subsection{Bi-local $\mathcal{N}=1$ Superconformal Generators}
\label{sec:bi-local superconformal generators}

In non-supersymmetric SYK models, it is useful to find eigenfunctions of the Casimir of the $SL(2)$ algebra in order to diagonalize the quadratic action because the Casimir commutes with the kernel of the quadratic action. Similarly, in the SUSY SYK model, it is important to consider generators of the $\mathcal{N}=1$ superconformal algebra given by
\begin{align}
\dP_a=&\partial_{\tau_a}\\
\dK_a=&\tau_a^2\partial_{\tau_a} + {1\over 3} \tau_a +\tau_a \theta_a\partial_{\theta_a}\label{def:SCT}\\
\dD_a=&\tau_a\partial_{\tau_a}  +{1\over 2} \theta_a \partial_{\theta_a} +{1\over 6}\\
\dQ_a=&\partial_{\theta_a} -\theta_a\partial_{\tau_a}\\
\dS_a=&\tau_a\partial_{\theta_a} - \tau_a \theta_a \partial_{\tau_a} - {1\over 3} \theta_a
\end{align}
where $a=1,2$. Note that the ${1\over 3}$ factors appear because the fermion has conformal dimension ${1\over 6}$. We define bi-local superconformal generator as follows.
\begin{equation}
    \mathcal{L}=\mathcal{L}_1+\mathcal{L}_2\hspace{1cm} (\; \mathcal{L}\in \{\dP,\dK,\dD,\dQ,\dS\}\;)
\end{equation}
which satisfy
\begin{alignat}{4}
&[\dP,\dK]=2 \dD&&\;,\quad \{\dQ,\dQ\}=-2\dP&&\;,\quad	[\dD,\dQ]=-{1\over 2} \dQ	&&\;,\quad	[\dP,\dQ]=0	\\
&[\dD,\dP]=-\dP&&\;,\quad \{\dQ,\dS\}=-2\dD&&\;,\quad	[\dD,\dS]={1\over 2} \dS	&&\;,\quad	[\dK,\dS]=0	\\
&[\dD,\dK]=\dK&&\;,\quad \{\dS,\dS\}=-2\dK&&\;,\quad	[\dK,\dQ]=-\dS	&&\;,\quad [\dP,\dS]=\dQ
\end{alignat}
The Casimir is given by
\begin{equation}
\mathcal{C}=\dD^2-{1\over2}(\dP\dK+\dK\dP)+{1\over 4} (\dS\dQ-\dQ \dS)=\dD^2-{1\over2} \dD-\dK\dP+{1\over 2} \dS\dQ
\end{equation}
Now, we will translate the generators as differential operators acting on superfields into supermatrices notation. Let us consider a superfield
\begin{equation}
    A^\mp(\tau_1,\tau_2)=A_0^\pm +\theta_1 A_1^\mp -A_2^\mp \theta_2 -\theta_1 A_3^\pm \theta_2= \begin{pmatrix}
    A_1^\mp & A_3^\pm\\
    A_0^\pm & A_2^\mp\\
    \end{pmatrix} \ .
\end{equation}
where we omit the bi-local time coordinates for a while. For example, one can consider the action of $\dK_1$ and $\dK_2$ in~\eqref{def:SCT} on the superfield $A^\mp$:
\begin{align}
    \dK_1 A^\mp=& \left(\tau_1^2\partial_{\tau_1} +{1\over 3}\tau_1\right)A_0^\pm +\theta_1\left(\tau_1^2\partial_{\tau_1} +{4\over 3}\tau_1\right)A_1^\mp \cr
    &-\left(\tau_1^2\partial_{\tau_1} +{1\over 3}\tau_1\right)A_2^\mp\theta_2 -\theta_1\left(\tau_1^2\partial_{\tau_1} +{4\over 3}\tau_1\right)A_3^\pm \theta_2 \\
    \dK_2 A^\mp=& \left(\tau_2^2\partial_{\tau_2} +{1\over 3}\tau_2\right)A_0^\pm +\theta_1\left(\tau_2^2\partial_{\tau_2} +{1\over 3}\tau_2\right)A_1^\mp \cr
    &-\left(\tau_2^2\partial_{\tau_2} +{4\over 3}\tau_2\right)A_2^\mp\theta_2 -\theta_1\left(\tau_2^2\partial_{\tau_2} +{4\over 3}\tau_2\right)A_3^\pm\theta_2
\end{align}
From the view point of super matrix, this can be written as
\begin{align}
\dK_1 A^\mp= \mK\mstar A^\mp \qquad,\qquad \dK_2 A^\mp= A^\mp \mstar \mK^\pst 
\end{align}
where $A^\pst$ is the composite operation of the parity transpose and supertranspose of a supermatrix $A$. Namely, the parity transpose of a supermatrix $A$ is defined by
\begin{equation}
    A=\begin{pmatrix}
    A_1 & A_3\\
    A_0 & A_2\\
    \end{pmatrix}\quad\Longrightarrow \quad A^\parity =\begin{pmatrix}
    A_2 & A_0\\
    A_3 & A_1\\
    \end{pmatrix}
\end{equation}
We define $A^\pst$ by 
\begin{equation}
    A^\pst=(A^\parity)^\st=\begin{pmatrix}
    A_2^t & (-1)^{|A|}A_3^t\\
    -(-1)^{|A|}A_0^t & A_1^t\\
    \end{pmatrix}
\end{equation}
Recall that $|A|$ denotes the parity of the supermatrix $A$. Repeating the same calculation for the other generators, we find that
\begin{equation}
    \mathcal{L}_1A=\mathbb{L}\mstar A\;,\;\; \mathcal{L}_2A=(-1)^{|\mathcal{L}|\cdot( |A|+1)} A\mstar \mathbb{L}^\pst  \hspace{0.5cm} (\; \mathcal{L}_a\in \{\dP_a,\cdots,\dS_a\}\;,\; \mathbb{L}\in \{\mP,\mK,\mD,\mQ,\mS\})
\end{equation}
where the supermatrices $\{\mP,\mK,\mD,\mQ,\mS\}$ are defined by
\begin{align}
    \mP\equiv& \begin{pmatrix}
    \partial_{\tau_1}\delta(\tau_1-\tau_2) & 0\\
    0 & \partial_{\tau_1}\delta(\tau_1-\tau_2) \\
    \end{pmatrix}\label{def:matrix P}\\
    \mK\equiv& \begin{pmatrix}
   (\tau_1^2\partial_{\tau_1}+{4\over 3}\tau_1)\delta(\tau_1-\tau_2) & 0\\
    0 & (\tau_1^2\partial_{\tau_1}+{1\over 3}\tau_1)\delta(\tau_1-\tau_2) \\
    \end{pmatrix}\label{def:matrix K}\\
    \mD\equiv& \begin{pmatrix}
     (\tau_1\partial_{\tau_1}+{2\over 3})\delta(\tau_1-\tau_2) & 0\\
    0 &  (\tau_1\partial_{\tau_1}+{1\over 6})\delta(\tau_1-\tau_2) \\
    \end{pmatrix}\label{def:matrix D}\\
    \mQ\equiv& \begin{pmatrix}
    0 & -\partial_{\tau_1} \delta(\tau_1-\tau_2)\\
    \delta(\tau_1-\tau_2) & 0 \\
    \end{pmatrix}\label{def:matrix Q}\\
    \mS\equiv& \begin{pmatrix}
    0 & (-\tau_1\partial_{\tau_1}-{1\over3}) \delta(\tau_1-\tau_2)\\
    \tau_1\delta(\tau_1-\tau_2) & 0 \\
    \end{pmatrix}\label{def:matrix S}
\end{align}
Note that $|\mathcal{L}|$ is the usual parity of the generator while $|A|$ is the parity as a supermatrix\footnote{Recall that parity of $A$ as a supermatrix is opposite to the ``usual parity'' of $A$ as a superfield.}.
Hence, the action of the  bi-local superconformal generator on the superfield can be represented as follows
\begin{equation}
    \mathcal{L}A=\mathbb{L}\mstar A+(-1)^{|\mathcal{L}|\cdot( |A|+1)} A\mstar \mathbb{L}^\pst  \hspace{1cm} (\; \mathcal{L}\in \{\dP,\dK,\dD,\dQ,\dS\}\;,\; \mathbb{L}\in \{\mP,\mK,\mD,\mQ,\mS\})\label{eq: matrix generator action}
\end{equation}
Note that the supermatrix generators are 
\begin{equation}
    |\mP|=|\mK|=|\mD|=0\qquad,\qquad |\mQ|=|\mS|=1
\end{equation}
Especially, $\mP$ and $\mQ$ satisfy
\begin{equation}
    \mP^\pst=-\mP\qquad,\qquad \mQ^\pst =\mQ\qquad,\qquad \mQ\mstar \mQ=\mP\ ,\label{eq: property of P and Q}
\end{equation}
and therefore, the action of $\dP$ and $\dQ$ are simply given by
\begin{equation}
\dP A= \mP \mstar A - A\mstar \mP\quad,\quad \dQ A= \mQ\mstar A + (-1)^{|A|+1}A \mstar \mQ \label{eq: property of P and Q 2}
\end{equation}

%
%

\subsection{Eigenfunctions of Superconformal Casimir}
\label{sec: eigenfunction of casimir}

In non-supersymmetric SYK model, it is natural to use new coordinates given by
\begin{equation}
t={1\over 2}(\tau_1+\tau_2)\quad,\quad z={1\over 2}(\tau_1-\tau_2)\label{eq:bi-local map}
\end{equation}
%
%
In fact, this is the simplest example of the bi-local map found in~\cite{Jevicki:2015pza,Koch:2014aqa,Koch:2014mxa,Koch:2010cy} for the duality between higher spin theory in AdS$_4$ and free vector model CFT$_3$. This bi-local map can be obtained by comparing the bi-local conformal generators for $O(N)/U(N)$ vector fields and and conformal generators for higher spin fields. But, the bi-local space of (non-supersymmetric) SYK model is so simple that we need not do such calculations\footnote{On the other hand, bi-local map of superspace might be non-trivial because there could be a mixing between $\tau_1,\tau_2$ and $\theta_1\theta_2$. For $\mathcal{N}=1$ SUSY SYK model, such a mixing does not seem to be natural.}. For the rest of Grassmannian odd coordinates, we do not transform, but we will relabel the coordinates by
\begin{align}
\theta_1= \zeta_0 \quad,&\quad \theta_2= \zeta_1 \cr
\partial_{\zeta_0}=\partial_{\theta_1} \quad,& \quad \partial_{\zeta_1}= \partial_{\theta_2}  \label{eq:bi-local map2}
\end{align}
Under this bi-local map, the superconformal generators can be expressed by
\begin{align}
    \dP=&\partial_t\\
    \dK=&(t^2+z^2)\partial_t+2tz\partial_z+t(\zeta_0\partial_{\zeta_0}+\zeta_1\partial_{\zeta_1})+z(\zeta_0\partial_{\zeta_0}-\zeta_1\partial_{\zeta_1})+{2\over 3}t\cr
    =&(-t^2+z^2)\partial_t+2t D+z(\zeta_0\partial_{\zeta_0}-\zeta_1\partial_{\zeta_1})\\
    \dD=&t\partial_t +z\partial_z +{1\over 2}\zeta_0\partial_{\zeta_0}+{1\over 2} \zeta_1\partial_{\zeta_1}+{1\over 3} \\
    \dQ=&-{1\over 2} \zeta_0(\partial_t+\partial_z)+{1\over 2} \zeta_1(-\partial_t+\partial_z)+\partial_{\zeta_0}+\partial_{\zeta_1}\\
    \dS=&(t+z)\partial_{\zeta_0}-(-t+z)\partial_{\zeta_1}-{1\over 2}\zeta_0(t+z)(\partial_t+\partial_z)-{1\over 2}\zeta_1(-t+z)(-\partial_t+\partial_z)\cr
    &-{1\over 3}(\zeta_0+\zeta_1)
\end{align}
and the corresponding Casimir operator is found to be
\begin{align}
    &\mathcal{C}=-{1\over 18} +{2\over 3} z\partial_z  +z^2(-\partial_t^2+\partial_z^2)-z\partial_t(\zeta_0 \partial_{\zeta_0}- \zeta_1\partial_{\zeta_1})+(z\partial_z+{1\over 6})(\zeta_0\partial_{\zeta_0}+\zeta_1\partial_{\zeta_1}) \cr
    &+{1\over 2} \zeta_0\zeta_1 \partial_{\zeta_1}\partial_{\zeta_0} -z\partial_{\zeta_1}\partial_{\zeta_0}-{1\over 6}\partial_z\zeta_0 \zeta_1  -{1\over 4z} (-z^2\partial_t^2+z^2\partial_z^2) \zeta_0\zeta_1 \cr
    &-({1\over 2}z\partial_z+{1\over 6})(\zeta_0\partial_{\zeta_1}+\zeta_1\partial_{\zeta_0}) -{1\over 2} z\partial_t (\zeta_0\partial_{\zeta_1}- \zeta_1 \partial_{\zeta_0})
\end{align}
Now, we will find (super-)eigenfunctions for the Casimir:
\begin{equation}
\mathcal{C}A(t,z,\zeta_0,\zeta_1)=\Lambda A(t,z,\zeta_0,\zeta_1)
\end{equation}
where the (super-)eigenfunction is given by
\begin{align}
A(t,z,\zeta_0,\zeta_1)=A_0(t,z)+\zeta_0 A_1(t,z)-A_2(t,z)\zeta_1 - \zeta_0 A_3(t,z)\zeta_1
\end{align}

First, we will focus on bosonic\footnote{Recall that bosonic bi-local superfield corresponds to Grassmannian odd supermatrix $A^-$.} eigenfunction, that is, $A_0$ is Grassmannian even. Then, acting with the Casimir on the eigenfunction, we have
\begin{align}
&\mathcal{C} A^-\cr
=&\left[ -{1\over 18}A_0+{2\over 3} z\partial_zA_0 + z^2(-\partial_t^2+\partial_z^2)A_0 +zA_3   \right] \cr
&+\zeta_0 \left[{1\over 9}A_1+{5\over 3} z\partial_zA_1+z^2(-\partial_t^2+\partial_z^2)A_1 - z\partial_t A_1    -{1\over 2} z\partial_z A_2-{1\over 2}z\partial_tA_2 -{1\over 6}A_2 \right]\cr
&- \left[ {1\over 9}A_2+{5\over 3} z\partial_zA_2+z^2(-\partial_t^2+\partial_z^2)A_2 + z\partial_t A_2   -{1\over 2}z\partial_z A_1+{1\over 2}z\partial_t A_1   -{1\over 6}A_1\right]\zeta_1 \cr
&- \zeta_0 \left[ {7\over 9}A_3+{8\over 3} z\partial_zA_3 + z^2(-\partial_t^2+\partial_z^2) A_3    +{1\over 6}\partial_z A_0+{1\over 4z} (-z^2\partial_t^2+z^2\partial_z^2)A_0 \right] \zeta_1
\end{align}
Note that $A_0$ (and, $A_1$) and $A_3$ ($A_2$, respectively) are mixed. For $A_0$ and $A_3$, we will use the following ansatz which is similar to non-supersymmetric SYK model~\cite{Polchinski:2016xgd,Jevicki:2016bwu}:
\begin{align}
A_0=&  e^{-iw t} z^{1\over 6}J_\nu(wz)\\
A_3=& a_3 e^{-iw t} z^{-{5\over 6}}J_\nu(wz)
\end{align}
%
%
We find that there are two solutions given by
\begin{equation}
a_3={1\over 2}\left({1\over 6}\pm \nu \right)
\end{equation}
and the corresponding eigenvalues are
\begin{equation}
\mathcal{C}A^-=\nu\left(\nu \pm {1\over 2}\right)A^-
\end{equation}
Since $\dQ $ commutes with the Casimir, $\dQ A^-$ is also an eigenfunction if $A^-$ is an eigenfunction. However, since that the parity of $\dQ A^-$ is opposite to $A$, $\dQ A^-$ is a fermionic eigenfunction. Furthermore, $A_0$ and $A_3$ components of the bosonic eigenvectors can determine the $A_1$ and $A_2$ components of the fermionic eigenfuction because of parity. This is also easily seen by the action of $\dQ$ on the (bosonic) eigenfunction:
\begin{align}
    \dQ A^-=&A_1+A_2 +\zeta_0\left(-{1\over 2}\partial_t A_0 - {1\over 2} \partial_z A_0+A_3\right) - \left({1\over 2} \partial_tA_0- {1\over 2}\partial_z A_0+A_3\right)\zeta_1 \cr
    &- \zeta_0 \left(-{1\over 2}\partial_t A_2-{1\over 2}\partial_z A_2+{1\over 2}\partial_t A_1-{1\over 2} \partial_z A_1\right)\zeta_1 \ .
\end{align}

In the same way, one can also find the $A_0$ and $A_3$ components of the fermionic eigenfunctions. \ie The action of the Casimir on the fermionic eigenfunction is 
\begin{align}
&\mathcal{C} A^+\cr
=&\left[-{1\over 18}A_0 +{2\over 3} z\partial_z A_0  + z^2(-\partial_t^2+\partial_z^2)A_0  -zA_3  \right] \cr
&+\zeta_0 \left[{1\over 9}A_1 +{5\over 3} z\partial_zA_1 + z^2(-\partial_t^2+\partial_z^2)A_1 - z\partial_t A_1 +{1\over 2}z\partial_zA_2+{1\over 2}z\partial_t A_2  +{1\over 6}A_2   \right]\cr
&- \left[ {1\over 9}A_2 +{5\over 3} z\partial_zA_2 + z^2(-\partial_t^2+\partial_z^2)A_2 + z\partial_t A_2  {1\over 2}z\partial_z A_1-{1\over 2}z\partial_t A_1+{1\over 6}A_1  \right]\zeta_1 \cr
&- \zeta_0 \left[ {7\over 9}A_3+{8\over 3} z\partial_zA_3 + z^2(-\partial_t^2+\partial_z^2) A_3 -{1\over 6}\partial_z A_0 -{1\over 4z}(-z^2\partial_t^2+z^2\partial_z^2)A_0 \right] \zeta_1\ .
\end{align}
Using an ansatz 
\begin{align}
A_0=&  e^{-iw t} z^{1\over 6}J_\nu(wz)\\
A_3=& a_3 e^{-iw t} z^{-{5\over 6}}J_\nu(wz)\ ,
\end{align}
we find that 
\begin{align}
a_3=&-{1\over 2}\left({1\over 6}\pm \nu\right)\\
\mathcal{C}A^+=&\nu\left(\nu \pm {1\over 2}\right)A^+
\end{align}
Now, $\dQ A^+$ gives $A_1$ and $A_2$ components of the bosonic eigenfunctions. \eg
\begin{align}
    \dQ A^+=&A_1-A_2 +\zeta_0\left(-{1\over 2}\partial_t A_0 - {1\over 2} \partial_z A_0-A_3\right) - \left(-{1\over 2} \partial_tA_0+ {1\over 2}\partial_z A_0+A_3\right)\zeta_1 \cr
    &- \zeta_0 \left(-{1\over 2}\partial_t A_2-{1\over 2}\partial_z A_2-{1\over 2}\partial_t A_1+{1\over 2} \partial_z A_1\right)\zeta_1 
\end{align}
We will also utilize the fermionic eigenfunctions of the Casimir in diagonalizing the quadratic action involved with fermi components in Section~\ref{sec: eigenfunctions of the quadratic action2}. We summarize all eigenfunctions in Appendix~\ref{app: casimir eigenfunction}.

\section{Diagonalization of the Quadratic Action}
\label{sec:diagonalization of quadratic action}

In this section, we will diagonalize the quadratic action in~\eqref{eq:quadratic action supermatrix}. For this, one can directly diagonalize the kernel as in~\cite{Polchinski:2016xgd} by using eigenfunctions for the Casimir found in the previous section because the classical solution (anti-)commutes with superconformal generators. \ie
\begin{equation}
 [\mathcal{L},\Psi_{cl}\}=\mathbb{L}\mstar \Psi_{cl}+ \Psi_{cl}\mstar \mathbb{L}^\pst \hspace{1cm} (\; \mathcal{L}\in \{\dP,\dK,\dD,\dQ,\dS\}\;,\; \mathbb{L}\in \{\mP,\mK,\mD,\mQ,\mS\})
\end{equation}
We give this direct diagonalization in Appendix~\ref{app: diagonalization} because they involve tedious integrations. Instead, we present the diagonalization in a pedagogical way based on an observation from the result of the direct evaluation.

The basic idea is to diagonalize separately  two terms in the quadratic action
\begin{equation}
S_{col}^{(2)}=-{1\over 2} \str \left(  \Psi_{cl}^{-1}\mstar \Phi\mstar \Psi_{cl}^{-1} \mstar \Phi + 2J\Phi\mstar [\Psi_{cl}\Phi]\right) \ .
\end{equation}
Indeed, we will see that the second term 
\begin{equation}
\str [\Phi\mstar [\Psi_{cl}\Phi]]
\end{equation}
is nothing but the inner product of two eignfunctions. In addition, in order to diagonalize the first term
\begin{equation}
\str [\Psi_{cl}^{-1}\mstar \Phi\mstar \Psi_{cl}^{-1} \mstar \Phi ] \ ,
\end{equation}
we will use a similar calculation as in~\cite{Jevicki:2016bwu}. That is, for each eigenfunction $u_{\nu w}$, we will find a function $\tilde{u}_{\nu w}$ such that
\begin{align}
    \Psi_{cl} \mstar \tilde{u}_{\nu w}\mstar \Psi_{cl}= g(\nu) u_{\nu w}
\end{align}
where $w$ is a frequency related to the eigenvalue of $\dP$, and $\nu$ is a representation of the superconformal algebra. In addition, $g(\nu)$ is a function of $\nu$, which will determine the spectrum of the SUSY SYK model.

\subsection{Eigenfunctions of the Quadratic Action: Bosonic Components}
\label{sec: eigenfunctions of the quadratic action}

\paragraph{Eigenfunctions:}

We begin with eigenfunction $u^1_{\nu w}$ of the superconformal Casimir in \eqref{eq:eigenfunction1}. This can be written as
\begin{equation}
e^{-iwt} z^{1\over 6 } J_\nu (|wz|)\begin{pmatrix}
    0 & -{\nu - {1\over 6} \over 2 \left|z\right|} \\
    \sgn(z) & 0 \\
    \end{pmatrix}\label{eq:eigenfunction temp1}
\end{equation}
Here, we demand that the eigenfunction $u^1_{\nu w}$ obeys the symmetry of the supermatrix of the $\mathcal{N}=1$ SYK model in~\eqref{eq: antisymmetry of supermatrix}. \ie
\begin{equation}
    \Jmatrix u_{\nu w}^\st \Jmatrix = u_{\nu w}\label{eq:anti-symmetry of eigenfunction}
\end{equation}
In general, we also have a second solution involved with $J_{-\nu}$ because the superconformal Casimir related to this eigenfunction is reduced to Bessel's differential equation. For the given $\nu$ and $w$, we have such an eigenfunction in the same representation in \eqref{eq:eigenfunction1--2} given by
\begin{equation}
e^{-iwt} z^{1\over 6 } J_{-\nu} (|wz|)\begin{pmatrix}
    0 & -{\nu - {1\over 6} \over 2 \left|z\right|} \\
    \sgn(z) & 0 \\
    \end{pmatrix}\label{eq:eigenfunction temp2}
\end{equation}
where we also demand the symmetry of the eigenfunction in~\eqref{eq:anti-symmetry of eigenfunction}. Hence, one has to find a relative coefficient of the eigenfunctions \eqref{eq:eigenfunction temp1} and \eqref{eq:eigenfunction temp2} to diagonalize the kernel of the quadratic action. This coefficient is usually determined by boundary condition. In particular, it is useful to think of the IR boundary condition (\ie $z\rightarrow \infty$).  
%
%
From the asymptotic behavior of the Bessel function, we have 
\begin{align}
    J_\nu(z)+\xi J_{-\nu}(z)\approx& \sqrt{2\over \pi z} \left(\cos{\pi\over 2}(\nu+1/2) + \xi \sin{\pi\over 2}(\nu+1/2) \right)\cos z\cr
    &+\sqrt{2\over \pi z}\left(\sin{\pi\over 2}(\nu+1/2) + \xi \cos{\pi\over 2}(\nu+1/2) \right)\sin z
\end{align}
where $\xi$ is a relative coefficient. In the non-supersymmetric SYK model, after direct diagonalization of the kernel, it turns out that the eigenfunction behaves like $z^{-{1\over 2}}\cos z$ in large $z$. In this section, we demand the generalized boundary condition thereof by brute force, but we also confirmed in Appendix~\ref{app: diagonalization} that this eigenfunctions indeed diagonalizes the quadratic action. In addition to the asymmptotic behavior $z^{-{1\over 2}} \cos z$, it would also possible to demand  $z^{-{1\over 2}} \sin z$ in large $z$. Hence, demanding those two boundary conditions, we generalize the function $Z_\nu(z)$ introduced in~\cite{Polchinski:2016xgd}:
\begin{equation}
    Z_\nu^\mp (z)\equiv J_\nu(z)+\xi_{\pm\nu}J_{-\nu}(z) 
\end{equation}
where $\xi_\nu$ is defined by
\begin{equation}
    \xi_\nu\equiv {\tan{\pi \nu\over 2}+ 1\over \tan{\pi \nu\over 2}-1 }
\end{equation}
Note that at large $z$, they behave as
\begin{equation}
    Z^-_\nu(z)\sim  {\cos z\over \sqrt{z}}\quad,\quad   Z^+_\nu(z)\sim {\sin z \over \sqrt{z}}
\end{equation}
Now, we will consider UV boundary condition ($z\rightarrow 0$). In \cite{Polchinski:2016xgd}, the Bessel's differential equation from the Casimir operator was interpreted as a Schordinger-like equation to claim that a real $\nu$ corresponds to a discrete bound state, and pure imaginary $\nu$'s are consist of continuum spectrum. Likewise, one can also expect that there are bound states for real $\nu$. Furthermore, we can also demand that the such eigenfunctions do not diverge as $z$ goes to zero. This gives a discrete series of possible $\nu$'s for each $Z_\nu^\mp$. \ie
\begin{align}
   Z^-_\nu(z)\quad:\quad \nu=2n+{3\over 2}\hspace{1cm} (n=0,1,2,\cdots)\\
   Z^+_\nu(z)\quad:\quad \nu=2n+{1\over 2}\hspace{1cm} (n=0,1,2,\cdots)
\end{align}
%
%
Now, since there are two independent linear combination of \eqref{eq:eigenfunction temp1} and \eqref{eq:eigenfunction temp2}, we have to determine which UV/IR boundary condition is possible for them. For this, we utilize the zero mode of the kernel involved with reparametrization. In non-supersymmetric SYK model, the zero mode can be evaluated \cite{Jevicki:2016ito} by
\begin{equation}
u_0(\tau_1,\tau_2)\equiv \left.{\delta  \Psi_{cl,f}(\tau_1,\tau_2) \over \delta f(\tau)}\right|_{f(\tau)=\tau}
\end{equation}
where $\Psi_{cl}$ is the large $N$ classical solution of non-supersymmetric SYK model, and $\Psi_{cl,f}$ is transformed classical solution by reparametrization $f(\tau)$. \ie
\begin{equation}
\Psi_{cl,f}=\left|f'(\tau_1)f'(\tau_2)\right|^{1\over q} \Psi_{cl}(f(\tau_1),f(\tau_2))
\end{equation}
In the SUSY SYK model, one can quickly obtain the zero mode from the classical solution in~\eqref{eq:classical solution q=3} by using the reparametrization instead of super-reparametrization. We found 
\begin{equation}
u_0\sim \begin{pmatrix}
0 & -{4\over 3|\tau_{12}|} \\
\sgn(\tau_{12}) & \\
\end{pmatrix}\label{eq:zero mode eigenfunction}
\end{equation}
It was already known that this zero mode corresponds to the eigenfunction $Z^-_{3\over 2} (z)$~\cite{Fu:2016vas}. On the other hand, we have two types of eigenfunctions \eqref{eq:eigenfunction1} or \eqref{eq:eigenfunction2}. For $\nu={1\over 2} $ or $\nu={3\over 2}$, we found that only \eqref{eq:eigenfunction1} with $\nu={3\over 2}$ can become the zero mode in \eqref{eq:zero mode eigenfunction}. Hence, we can deduce that \eqref{eq:eigenfunction1} satisfy the boundary condition of $Z^-_\nu$, and therefore, we can write the eigenfunction as
\begin{equation}
    u^1_{\nu w}(t,z)=e^{-iwt} |J z|^{1\over 6}Z^-_\nu(|wz|)\begin{pmatrix}
    0 & -{\nu -{1\over 6}\over 2|z|} \\
    \sgn(z) & 0 \\
    \end{pmatrix}\label{eq: u1 eigenfunction}
\end{equation}
or equivalently,
\begin{equation}
    u^1_{\nu w}(\tau_1,\tau_2)={e^{-{iw\over 2} (\tau_1+\tau_2)}\over \sqrt{8\pi}}\left|J{\tau_1-\tau_2\over 2}\right|^{1\over 6} Z^-_\nu(|{w\over 2}(\tau_1-\tau_2)|)\begin{pmatrix}
    0 & -{\nu - {1\over 6} \over 2 \left|{1\over 2}(\tau_1-\tau_2)\right|} \\
    \sgn(\tau_1-\tau_2) & 0 \\
    \end{pmatrix}
\end{equation}
where the representation $\nu$ can be either a pure imaginary continuum value or a discrete real value for UV boundary condition as in~\cite{Polchinski:2016xgd}. \ie
\begin{align}
    \nu=&{3\over 2}+2n\hspace{1cm} (n=0,1,2,\cdots)\\
    \nu=&ir\hspace{1cm} (r\geqq 0)
\end{align}
For the other UV/IR boundary condition, we have the eigenfunction \eqref{eq:eigenfunction2--1} corresponding to $Z^+_\nu$:
\begin{equation}
    u^2_{\nu w}(t,z)={e^{-iwt}\over \sqrt{8\pi}} |J z|^{1\over 6}Z^+_\nu(|wz|)\begin{pmatrix}
    0 & {\nu + {1\over 6}\over 2|z|} \\
    \sgn(z) & 0 \\
    \end{pmatrix}\label{eq: u2 eigenfunction}
\end{equation}
or equivalently, 
\begin{equation}
    u^2_{\nu w}(\tau_1,\tau_2)={e^{-{iw\over 2} (\tau_1+\tau_2)}\over\sqrt{8\pi}}\left|J{\tau_1-\tau_2\over 2}\right|^{1\over 6} Z^+_\nu(|{w\over 2}(\tau_1-\tau_2)|)\begin{pmatrix}
    0 & {\nu + {1\over 6} \over 2 \left|{1\over 2}(\tau_1-\tau_2)\right|} \\
    \sgn(\tau_1-\tau_2) & 0 \\
    \end{pmatrix}
\end{equation}
where we also demanded the symmetry of eigenfunctions in \eqref{eq:anti-symmetry of eigenfunction}, and the representation $\nu$'s are
\begin{align}
    \nu=&{1\over 2}+2n\hspace{1cm} (n=0,1,2,\cdots)\\
    \nu=&ir\hspace{1cm} (r\in \mathbb{R})
\end{align}

\paragraph{Diagonalization of the second term:}

It is useful to find orthogonality of the functions $Z^\mp_\nu$'s because the second term in the quadratic action in~\eqref{eq:quadratic action supermatrix} is, in fact, reduced to an inner product of $Z^\mp_{\nu}$'s. \ie
\begin{equation}
\str\left(u_{\nu,w}\mstar [\Psi_{cl} u_{\nu',w'}]\right)\sim \delta(w+w') \int_0^\infty {dz\over z} Z_\nu^\alpha (z) Z_{\nu'}^{\alpha'} (z)\label{eq:the first term in quadratic action}
\end{equation}
where $\alpha, \alpha'=\mp$. First, it is easy to see that $Z^-_\nu$ is orthogonal to $Z^+_\nu$ because they have different eigenvalues for Casimir. By a similar analysis to~\cite{Polchinski:2016xgd}, we found that
\begin{equation}
   \int_0^\infty  { dz \over |z|}  \overline{Z^\alpha_\nu(|wz|)} Z^{\alpha}_{\nu'}(|w'z|)=\delta_{\alpha \alpha'} N_\nu \delta(\nu-\nu')\label{eq:orthogonality of z}
\end{equation}
where
\begin{equation}
    N_\nu=\begin{cases}
    \quad {1\over 2\nu} & \quad \left(\nu={3\over 2}+2n\;\;\mbox{for $Z^-$, or}\;\;\nu={1\over 2}+2n \;\;\mbox{for $Z^+$}\quad(n=0,1,2,\cdots)\right)\\
    \quad {2 \sin \pi \nu\over \nu} & \quad \left(\mbox{for}\quad \nu=ir \quad(r\in \mathbb{R})\right)
    \end{cases}
\end{equation}
For real $\nu$, $Z^\mp_\nu$ is a real function so that we can immediately see that \eqref{eq:the first term in quadratic action} is diagonalized. On the other hand, for pure imaginary value $\nu=ir$, the complex conjugate of the function $Z_\nu^\mp $ can be written as
\begin{equation}
    \overline{Z_\nu^\mp }=J_{-\nu}(z) +\xi_{\mp \nu} J_\nu (z)=\xi_{\mp \nu} \left[ J_\nu(z) +\xi_{\pm \nu} J_{-\nu}(z) \right]=\xi_{\mp \nu} Z_\nu^\mp (z)
\end{equation}
where we used a useful identity for $\xi_\nu$:
\begin{equation}
    \xi_{-\nu}\xi_\nu=1
\end{equation}
Hence, we have
\begin{align}
    \int_0^\infty  { dz \over |z|}  Z^\mp_{ir}(|wz|) Z^\mp_{ir'}(|w'z|)=\widetilde{N}^\mp_{ir} \delta(r-r')\hspace{1cm} ( \widetilde{N}^\mp_{ir} \equiv \xi_{\pm ir } N_{ir})
\end{align}
and, \eqref{eq:the first term in quadratic action} is also diagonalized. We emphasize that \eqref{eq:the first term in quadratic action} leads to an induced inner product for the supermatrix formulation. \ie 
\begin{equation}
\langle u_{\nu,w}, u_{\nu', w'}\rangle \equiv\str\left(u_{\nu,w} \mstar [\Psi_{cl} u_{\nu',w'}]\right)\label{def:induced innner product}
\end{equation}

\paragraph{Diagonalization of the first term:}

Next, let us consider the first term in \eqref{eq:quadratic action supermatrix}. To diagonalize it, for each $u_{\nu w}$, we will find a function $\tilde{u}_{\nu w}$ such that 
\begin{align}
    \Psi_{cl} \mstar \tilde{u}_{\nu w}\mstar \Psi_{cl}= g(\nu) u_{\nu w}\label{eq: tilde u def}
\end{align}
where $g(\nu)$ is a function of $\nu$. In Appendix~\ref{app: diagonalization}, one can directly find $\tilde{u}$ for each $u^1_{\nu w}$ and $u^2_{\nu w}$. But, in this section, we present a new method to find $\tilde{u}$.

Suppose that there exist $\tilde{u}_{\nu w}$ to satisfy \eqref{eq: tilde u def}. Then, the first term in \eqref{eq:quadratic action supermatrix} becomes
\begin{equation}
\str(u_{\nu' w'} \mstar \Psi_{cl}^{-1}\mstar u_{\nu w} \mstar \Psi_{cl}^{-1})={1\over g(\nu)} \str (u_{\nu' w'} \mstar \tilde{u}_{\nu w})\label{eq:diagonalization of the first term}
\end{equation}
one may find a function $\doubletilde{u}_{\nu w}$ such that 
\begin{equation}
\tilde{u}_{\nu w}(\tau_1,\tau_2)= [\Psi_{cl}\doubletilde{u}_{\nu w}](\tau_1,\tau_2)
\end{equation}
where the product on the RHS is the usual product of superfields. Then, \eqref{eq:diagonalization of the first term} becomes
\begin{equation}
\str(u_{\nu' w'} \mstar \Psi_{cl}\mstar u_{\nu w} \mstar \Psi_{cl})=\langle u_{\nu' w'} , \doubletilde{u}_{\nu w}\rangle
\end{equation}
where $\langle \; \cdot\;,\; \cdot\;\rangle  $ is the induced inner product of supermatrix defined in \eqref{def:induced innner product}. Hence, if the first term is diagonalized by $u_{\nu w}$, we should have
\begin{equation}
\tilde{u}_{\nu w}(\tau_1,\tau_2)\sim  [\Psi_{cl} u_{\nu w}](\tau_1,\tau_2)\label{eq: tilde solution}
\end{equation}
Of course, this is confirmed by direct calculation for $q=3$ case as well as general $q$ case where $\Psi_{cl}$ on the RHS of \eqref{eq: tilde solution} and \eqref{def:induced innner product} is replaced by $\Psi_{cl}^{q-2}$. The remaining calculation is to fix the coefficient and the function $g(\nu)$ where one cannot avoid evaluating integrations. We found that
\begin{align}
    \tilde{u}^1_{\nu w}(\tau_1,\tau_2)=&
    -{2A\over \sqrt{8\pi}} J {e^{-{iw\over 2} (\tau_1+\tau_2)}\over \left|{J\over 2}(\tau_1-\tau_2)\right|^{1\over 6}} Z^-_\nu(|{w\over 2}(\tau_1-\tau_2)|)\begin{pmatrix}
    0 & -{\nu +{1\over 6} \over 2\left|{1\over 2}(\tau_1-\tau_2)\right|}\sgn(\tau_1-\tau_2) \\
    1 & 0 \\
    \end{pmatrix}\label{eq:tilde u sol1} \\
    \tilde{u}^2_{\nu w}(\tau_1,\tau_2)=&{2A\over \sqrt{8\pi}} J {e^{-{iw\over 2} (\tau_1+\tau_2)}\over \left|{J\over 2}(\tau_1-\tau_2)\right|^{1\over 6}} Z^+_\nu(|{w\over 2}(\tau_1-\tau_2)|)\begin{pmatrix}
    0 & {\nu -{1\over 6} \over 2 \left|{1\over 2}(\tau_1-\tau_2)\right|}\sgn(\tau_1-\tau_2) \\
    1 & 0 \\
    \end{pmatrix}\label{eq:tilde u sol2}
\end{align}
where $A=\left( {1\over 4\sqrt{3} \pi }\right)^{1\over 3}$ and
\begin{align}
    g_1(\nu)=&-2^{-{1\over 3} }{\Gamma\left({5\over 3}\right) \Gamma\left({5\over 12}-{\nu\over 2}\right) \Gamma\left({5\over 12}+{\nu\over 2}\right)\over \Gamma\left({4\over 3}\right)\Gamma\left({13\over 12}-{\nu\over 2}\right)\Gamma\left({1\over 12}+{\nu\over 2}\right) }\label{eq:spectrum eq1}\\
    g_2(\nu)=&-2^{-{1\over 3} }{\Gamma\left({5\over 3}\right) \Gamma\left({5\over 12}-{\nu\over 2}\right) \Gamma\left({5\over 12}+{\nu\over 2}\right)\over \Gamma\left({1\over 2}\right)\Gamma\left({1\over 3}-{h\over 2}\right)\Gamma\left({13\over 12}+{h\over 2}\right) }\label{eq:spectrum eq2}
\end{align}
which agrees with~\cite{Fu:2016vas}. Note that $\tilde{u}_{\nu w}$'s in \eqref{eq:tilde u sol1} and \eqref{eq:tilde u sol2} have different symmetry from $u_{\nu w}$. \ie
\begin{equation}
    \Jmatrix\mstar\tilde{u}_{\nu w}^\st\mstar\Jmatrix=- \tilde{u}_{\nu w}\label{eq: u tilde sym}
\end{equation}
This can be easily seen from the definition of $\tilde{u}_{\nu w}$ in \eqref{eq: tilde u def}:
\begin{equation}
    \Jmatrix (\Psi_{cl} \mstar \tilde{u}_{\nu w}\mstar \Psi_{cl})^\st\mstar \Jmatrix = -\Psi_{cl} \mstar \Jmatrix\mstar\tilde{u}_{\nu w}^\st\mstar\Jmatrix\mstar \Psi_{cl}=\Psi_{cl} \mstar \tilde{u}_{\nu w}\mstar \Psi_{cl}
\end{equation}

Now, we expand the fluctuation $\Phi$ in \eqref{eq:quadratic action supermatrix} in terms of $u^1_{\nu w}$ and $u^2_{\nu w}$:
\begin{equation}
\Phi= \sum_{w}\left[\sum_{\substack{\nu=2n+{3\over 2}\\ n=0,1,\cdots}}\oscA^1_{\nu w}  u^1_{\nu w} +\sum_{\substack{\nu=2n+{1\over 2}\\ n=0,1,\cdots}}\oscA^2_{\nu w}  u^2_{\nu w} + \sum_{\substack{\nu=ir\\ r\geqq 0}}  \left(\oscA^1_{\nu w}  u^1_{\nu w}+\oscA^2_{\nu w}  u^2_{\nu w}\right)\right]
\end{equation}
Note that the reality condition of the component fields leads to 
\begin{equation}
\overline{\Psi}=-\Psi\label{eq:reality condition}
\end{equation}
which imposes the following constraint.
\begin{alignat}{2}
&\overline{\oscA^1_{\nu w}}= -\oscA^1_{\nu -w}&& \mbox{for}\;\;\nu=2n+{3\over 2} \quad(n=0,1,2,\cdots)\\
&\overline{\oscA^2_{\nu w}}= -\oscA^2_{\nu -w}&& \mbox{for}\;\;\nu=2n+{1\over 2} \quad(n=0,1,2,\cdots)\\
&\overline{\oscA^1_{\nu w}}= -\xi_\nu\oscA^1_{\nu -w}&& \mbox{for}\;\;\nu=ir \quad(r\geqq 0)\\
&\overline{\oscA^2_{\nu w}}= -\xi_{-\nu}\oscA^2_{\nu -w}&& \mbox{for}\;\;\nu=ir \quad(r\geqq 0)
\end{alignat}
Then, we found that the quadratic action in \eqref{eq:quadratic action supermatrix} can be written as
\begin{align}
   S^{(2)}_{col}=& { 2J \over 2^{2\over 3} 3^{1\over 6} \pi^{1\over 3} }\sum_{w\geqq0}\;\;\sum_{\nu= 2n+{3\over 2}\;\text{{\tiny or}}\; \nu=ir } \nu N_\nu{1-g_1(\nu)\over g_1(\nu)}\left|\oscA^1_{\nu w}\right|^2\cr
   & +{ 2J \over 2^{2\over 3} 3^{1\over 6} \pi^{1\over 3} }\sum_{w\geqq0}\;\;\sum_{\nu= 2n+{1\over 2} \;\text{{\tiny or}}\; \nu=ir  } \nu N_\nu{1-g_2(\nu)\over g_2(\nu)}\left|\oscA^2_{\nu w}\right|^2
\end{align}
where we absorbed the factor $\xi_{\pm\nu}$ in the normalization $\widetilde{N}^\pm_{\nu} =\xi_{\pm \nu} N_\nu$ into the reality condition. This leads to two-point function of bi-local collective superfields (or, invariant four-point function of fundamental superfield). The summation over $\nu=ir$ can be understood as a contour integral along the imaginary axis. Repeating the same procedure in~\cite{Jevicki:2016bwu,Maldacena:2016hyu}, one can expect that the contour integral will pick up simples poles comes from $1-g_1(\nu) $ and $1-g_2(\nu)$ and the residues from other simple poles will cancel with the contribution from discrete series of $\nu$. Hence, the half of the spectrum of the $\mathcal{N}=1$ SUSY SYK model is given by two equations
\begin{equation}
g_1(\nu)=1\quad,\quad g_2(\nu)=1
\end{equation}
which was shown in\cite{Fu:2016vas}.

One can also diagonalize the quadratic action with the following fermionic eigenfunctions:
\begin{align}
    u^3_{\nu w}(t,z)=&e^{-iwt} |J z|^{1\over 6}Z^-_\nu(|wz|)\begin{pmatrix}
    0 & \oscB_{\nu w}{\nu -{1\over 6}\over 2|z|} \\
    \oscB_{\nu w}\sgn(z) & 0 \\
    \end{pmatrix}\\
    u^4_{\nu w}(t,z)=&{e^{-iwt}\over \sqrt{8\pi}} |J z|^{1\over 6}Z^+_\nu(|wz|)\begin{pmatrix}
    0 & -\oscB_{\nu w}{\nu + {1\over 6}\over 2|z|} \\
    \oscB_{\nu w}\sgn(z) & 0 \\
    \end{pmatrix}
\end{align}
where $\oscB_{\nu w}$ is a Grassmannian odd constant. Comparing to $u^1_{\nu w}$ and $u^2_{\nu w}$ in \eqref{eq: u1 eigenfunction} and \eqref{eq: u2 eigenfunction}, one can see that the only difference is the sign of $\theta_1\theta_2$ components. Moreover, because $\oscB_{\nu w}$ is Grassmannian odd, one can ends up with the same calculations as those in bosonic Grassmannian eigenfunctions except for an overall minus sign.

\subsection{Eigenfunctions of the Quadratic Action: Fermionic Components}
\label{sec: eigenfunctions of the quadratic action2}

After obtaining the bosonic eigenfunctions and the corresponding eigenvalues for the kernel, the diagonalization by fermionic components of bosonic eigenfunction is straightforward because of supersymmetry. In this section, we work out this diagonalization in detail. Also, we double-checked a part of the diagonalization by direct calculation in Appendix~\ref{app: diagonalization}.

We claim that $\dQ u_{\nu w}^a$ $(a=3,4)$ diagonalize the quadratic action with the same eigenvalue as $u_{\nu w}^a$. First, note that the classical solution $\Psi_{cl}$ is annihilated by the bi-local supercharge $\dQ$ which we have discussed in \eqref{sec:bi-local superconformal generators}
\begin{equation}
    \dQ \Psi_{cl}=\mQ \mstar \Psi_{cl} +\Psi_{cl}\mstar \mQ=0\label{eq: Q acts on classical solution}
\end{equation}
where $\mQ$ is defined in \eqref{def:matrix Q}. 

Now, we will find an analogous identity to \eqref{eq: tilde u def}. We will act with $\dQ \mB\mstar $ on the both sides of \eqref{eq: tilde u def} where $\mB$ is a constant Grassmannian odd supermatrix defined by
\begin{equation}
\mB=\begin{pmatrix}
\oscB & 0 \\
0 & \oscB\\
\end{pmatrix}\hspace{1cm} \oscB\;\;:\;\;\mbox{Grassmannian odd constant.}
\end{equation}
Note that the supermatrix $\mB$ commutes with $\mQ, \mP$ and $\Psi_{cl}$. Using \eqref{eq: property of P and Q 2} and \eqref{eq: Q acts on classical solution}, it becomes
\begin{align}
    g\mathcal{Q}(\mB\mstar u )=&\dQ(\Psi_{cl}\mstar \mB\mstar  \tilde{u} \mstar \Psi_{cl})=  \mQ\mstar \Psi_{cl}\mstar \mB\mstar\tilde{u} \mstar \Psi_{cl} - \Psi_{cl}\mstar  \mB\mstar\tilde{u} \mstar \Psi_{cl}\mstar \mQ\cr
       =&-\left(\Psi_{cl}\mstar \mQ\mstar\mB \mstar \tilde{u} \mstar \Psi_{cl}-\Psi_{cl}\mstar \mB\mstar \tilde{u}\mstar \mQ \mstar \Psi_{cl}\right)=-\Psi_{cl}\mstar [\mathcal{Q}(\mB\mstar \tilde{u})] \mstar \Psi_{cl}
\end{align}
where we omit $\nu$ and $w$. Hence, for the given $u_{\nu w}$, $\dQ(\mB\mstar  u_{\nu w})$ and $\dQ(\mB\mstar \tilde{u}_{\nu w})$ satisfy \eqref{eq: tilde u def} with the same $g(\nu)$, but with an additional minus sign. \ie
\begin{equation}
	g(\nu) \mathcal{Q}( \mB\mstar u_{\nu w})= -\Psi_{cl}\mstar \dQ(\mB\mstar  \tilde{u}_{\nu w}) \mstar \Psi_{cl}
\end{equation}
This simplify the first term in \eqref{eq:quadratic action supermatrix}, and we need to evaluate $\str[\dQ(\mB\mstar u)\mstar  \dQ (\mB \mstar \tilde{u})]$. Using \eqref{eq: property of P and Q} and \eqref{eq: property of P and Q 2}, we have
\begin{align}
    &\str[\dQ(\mB\mstar u)\mstar \dQ( \mB\mstar \tilde{u})]=\str\left[(\mQ\mstar\mB\mstar u - \mB\mstar u\mstar \mQ)\mstar ( \mQ\mstar \mB\mstar \tilde{u}- \mB\mstar \tilde{u}\mstar \mQ)\right]\cr
    =& -\str[-\mQ \mstar \mQ \mstar (\mB\mstar  u) \mstar (\mB\mstar\tilde{u})+(\mB \mstar u) \mstar \mQ \mstar \mQ \mstar (\mB\mstar \tilde{u}) ]\cr
    &+\str[\mQ \mstar(\mB\mstar u)  \mstar \mQ \mstar (\mB\mstar\tilde{u}) -   \mQ \mstar (\mB\mstar u) \mstar \mQ \mstar (\mB\mstar \tilde{u}) ]\cr
    =& -\str [ \mP \mstar (\mB\mstar u)\mstar (\mB\mstar \tilde{u}) - (\mB\mstar u) \mstar \mP\mstar ( \mB\mstar \tilde{u})]=-\str[(\dP(\mB\mstar   u))\mstar (\mB\mstar \tilde{u})]
\end{align}
where we used the following property of the supertrace in the second line 
\begin{equation}
    \str (XY)=(-1)^{|X|\cdot|Y|}\str (YX)
\end{equation}
Therefore, the first term in the quadratic action can be written as
\begin{equation}
    -{1\over 2} \str[ \dQ(\mB_{\nu' w'}\mstar u_{\nu' w'})\mstar \Psi_{cl}^{-1} \mstar \dQ(\mB_{\nu w} \mstar  u_{\nu w}) \mstar \Psi_{cl}^{-1}  ]
    =-{1\over 2g(\nu)} \str [(\dP(\mB_{\nu' w'}\mstar u_{\nu' w'}))\mstar \mB_{\nu w} \mstar \tilde{u}_{\nu w}]\label{eq:fermi diagonal result1}
\end{equation}
and, this corresponds to diagonalization of Grassmannian odd eigenfunctions in the previous section. 

In a similar way, one can also show the $\dQ u$ will diagonalize the second term of \eqref{eq:quadratic action supermatrix}. For this, we need to move the differential operator $\dQ$ by using integration by parts in the superspace integration. But, in the supermatrix formulation, this is nothing but property of supertrace. \eg
\begin{align}
    &\str[ (\dQ X)\mstar Y]=\str[\mQ \mstar X  \mstar Y]+(-1)^{|X|+1}\str[  X \mstar \mQ \mstar Y]\cr
    =&(-1)^{|X|+1} \str [X \mstar (\dQ Y)]
\end{align}
Thus, the inner product of two $\dQ (\mB \mstar u)$ is given by
\begin{align}
     &\langle \dQ (\mB\mstar u), \dQ (\mB\mstar u) \rangle  =-\str[\mB\mstar u\mstar \dQ (\Psi_{cl} \dQ(\mB\mstar u)) ]\cr
     =&-\str[\mB\mstar u\mstar  (\Psi_{cl}\dQ^2(\mB \mstar  u)) ]=\str[\mB\mstar u\mstar  (\Psi_{cl}\dP(\mB \mstar  u)) ]\cr
    =& \int d\tau_1 d\theta_1 d\tau_2 d\theta_2\; \Psi_{cl}(\tau_1,\theta_1;\tau_2,\theta_2) [\mB\mstar u](\tau_1,\theta_1;\tau_2,\theta_2) [\dP (\mB \mstar u)](\tau_1,\theta_1;\tau_2,\theta_2)\label{eq:fermi diagonal result1}
\end{align}
In the same way as before, we expand the fluctuation $\Phi$ in terms of $\dQ (\mB^1_{\nu w} \star u^3_{\nu w})$ and $\dQ (\mB^1_{\nu w} \star u^4_{\nu w})$, and the diagonalization is exactly the same as those of $u^3_{\nu w}$ and $u^4_{\nu w}$ which we shortly discussed before.

\section{$\mathcal{N}=2$ Supersymmetric SYK Model}
\label{sec:n=2 syk model}

In this section, we will generalize $\mathcal{N}=1$ bi-local collective superfield theory to $\mathcal{N}=2$ case. 

\subsection{Bi-local Chira/Anti-chiral Superspace, Superfield and Supermatrix}
\label{sec:bi-local superspace n=2}

We begin with the bi-local superspace for $\mathcal{N}=2$ SUSY vector models. At first glance, it seems that we have a larger Grassmannian space because there are two Grassmannian coordinates $\theta$ and $\bar{\theta}$. However, since we will focus on the chiral or anti-chiral superfields, the construction is almost the same as for $\mathcal{N}=1$ case. First, let us focus on superfield $A$ which is chiral with respect to the first superspace and anti-chiral in the second superspace:
\begin{equation}
\overline{\sD}_1A(\tau_1,\theta_1,\bar{\theta}_1;\tau_2,\bar{\theta}_2,\theta_2)=\sD_2A(\tau_1,\theta_1,\bar{\theta}_1;\tau_2,\bar{\theta}_2,\theta_2)=0
\end{equation}
where the superderivatives are given by
\begin{equation}
\sD\equiv \partial_\theta + \bar{\theta} \partial_\tau\;\;,\quad \bar{\sD}\equiv \partial_{\bar{\theta}} + \theta \partial_\tau
\end{equation}
Hence, the superfield $A$ depends only on $(\sigma_1,\theta_1;\bar{\sigma}_2,\bar{\theta}_2)$ where
\begin{equation}
\sigma\equiv \tau+\theta\bar{\theta}\quad,\quad \bar{\sigma}\equiv \tau-\theta\bar{\theta}
\end{equation}
and, one can expand the superfield $A$ as follows.
\begin{align}
A(\sigma_1,\theta_1;\bar{\sigma}_2,\bar{\theta}_2)=&A_0(\sigma_1,\bar{\sigma}_2)+\theta_1 A_1(\tau_1,\bar{\sigma}_2)- A_2 (\sigma_1,\tau_2) \bar{\theta}_2 - \theta_1 A_3(\tau_1,\tau_2)\bar{\theta}_2\\
=&A_0(\sigma_1,\bar{\sigma}_2)+\theta_1 A_1(\sigma_1,\bar{\sigma}_2)- A_2 (\sigma_1,\bar{\sigma}_2) \bar{\theta}_2 - \theta_1 A_3(\sigma_1,\bar{\sigma}_2)\bar{\theta}_2
\end{align}
This bi-local superfield naturally appears in the $U(N)$ vector models because chiral superfields and anti-chiral superfields transform in the fundamental and anti-fundamental representations of $U(N)$, respectively so that they form a $U(N)$ invariant bi-local field. Hence, it is natural to construct the following bi-local superspace for such bi-local $U(N)$ superfields.
\begin{equation}
(\sigma_1,\theta_1;\bar{\sigma}_2,\bar{\theta}_2)
\end{equation}

Now, we will define a star product in this bi-local superspace. However, it is difficult to construct the consistent star product of two chiral/anti-chiral bi-locals because the first and the second superspace have opposite chirality. Hence, we also introduce conjugate anti-chiral/chiral bi-local super field:
\begin{equation}
\bar{B}(\bar{\sigma}_1,\bar{\theta}_1;\sigma_2,\theta_2)= \bar{B}_0(\bar{\sigma}_1,\sigma_2)+ \bar{\theta}_1 \bar{B}_1(\bar{\sigma}_1,\sigma_2)- \bar{B}_2(\bar{\sigma}_1,\sigma_2)\theta_2 -   \bar{\theta}_1\bar{B}_3(\bar{\sigma}_1,\sigma_2) \theta_2
\end{equation}
We found that a consistent star product between $A(\sigma_1,\theta_1;\bar{\sigma}_2,\bar{\theta}_2)$ and $\bar{B}(\bar{\sigma}_1,\bar{\theta}_1;\sigma_2,\theta_2)$ is given by
\begin{equation}
A\bar{\mstar} \bar{B} \equiv\int A(\sigma_1,\theta_1;\bar{\sigma}_3,\bar{\theta}_2)d\tau_3 d\bar{\theta}_3  \bar{B}(\bar{\sigma}_3,\bar{\theta}_3;\sigma_2,\theta_2)
\end{equation}
which was already recognized in~\cite{Fu:2016vas} to analyze the Schwinger-Dyson equation. Similarly, we also define
\begin{equation}
\bar{B}\mstar A  \equiv\int  \bar{B}(\bar{\sigma}_1,\bar{\theta}_1;\sigma_3,\theta_3) d\tau_3 d\theta_3  A(\sigma_3,\theta_3;\bar{\sigma}_2,\bar{\theta}_2)
\end{equation}
Note that $A\bar{\mstar} \bar{B}$ is a chiral/chiral superfield while $\bar{B}\mstar A $ is an anti-chiral/anti-chiral superfield. As in $\mathcal{N}=1$ case, the punchline is that the supermatrix formulation drastically simplifies this complicated star product in the bi-local superspace into matrix multiplication. First, we represent the bi-local superfields $A$ and $\bar{B}$ as the following supermatrix.
\begin{align}
A(\sigma_1,\theta_1;\bar{\sigma}_2,\bar{\theta}_2)=&A_0(\sigma_1,\bar{\sigma}_2)+\theta_1 A_1(\sigma_1,\bar{\sigma}_2)- A_2 (\sigma_1,\bar{\sigma}_2) \bar{\theta}_2 - \theta_1 A_3(\sigma_1,\bar{\sigma}_2)\bar{\theta}_2\cr
\equiv&\begin{pmatrix}
A_1(\sigma_1,\bar{\sigma}_2)& A_3(\sigma_1,\bar{\sigma}_2) \\
A_0(\sigma_1,\bar{\sigma}_2) & A_2(\sigma_1,\bar{\sigma}_2) \\
\end{pmatrix}\\
\bar{B}(\bar{\sigma}_1,\bar{\theta}_1;\sigma_2,\theta_2)=& \bar{B}_0(\bar{\sigma}_1,\sigma_2)+ \bar{\theta}_1 \bar{B}_1(\bar{\sigma}_1,\sigma_2)- \bar{B}_2(\bar{\sigma}_1,\sigma_2)\theta_2 -   \bar{\theta}_1\bar{B}_3(\bar{\sigma}_1,\sigma_2) \theta_2\cr
\equiv&\begin{pmatrix}
\bar{B}_1(\bar{\sigma}_1,\sigma_2)  & \bar{B}_3 (\bar{\sigma}_1,\sigma_2) \\
\bar{B}_0(\bar{\sigma}_1,\sigma_2)  & \bar{B}_2 (\bar{\sigma}_1,\sigma_2) \\
\end{pmatrix}
\end{align}
Then, one can show that the star product of superfields becomes the following matrix product:
\begin{equation}
A\bar{\mstar} \bar{B}=\begin{pmatrix}
A_1 & A_3\\
A_0 & A_2\\
\end{pmatrix}\bar{\mstar}\begin{pmatrix}
\bar{B}_1 & \bar{B}_3\\
\bar{B}_0 & \bar{B}_2\\
\end{pmatrix}\quad,\quad \bar{B}\star A=\begin{pmatrix}
\bar{B}_1 & \bar{B}_3\\
\bar{B}_0 & \bar{B}_2\\
\end{pmatrix} \mstar\begin{pmatrix}
A_1 & A_3\\
A_0 & A_2\\
\end{pmatrix}
\end{equation}
These matrix products $\mstar$ and $\bar{\mstar}$ are a combination of the usual matrix product and star product~$\star$ in bi-local time space $(\tau_1,\tau_2)$ like the $\mathcal{N}=1$ case:
\begin{equation}
\begin{pmatrix}
A_1 & A_3\\
A_0 & A_2\\
\end{pmatrix}\bar{\mstar}\begin{pmatrix}
\bar{B}_1 & \bar{B}_3\\
\bar{B}_0 & \bar{B}_2\\
\end{pmatrix}=\begin{pmatrix}
(A_1\star \bar{B}_1+A_3\star \bar{B}_0 )   &\quad (A_1\star \bar{B}_3+ A_3\star \bar{B}_2)  \\
(A_0\star \bar{B}_1+ A_2\star \bar{B}_0) &\quad (A_0\star \bar{B}_3 + A_2\star \bar{B}_2)  \\
\end{pmatrix}
\end{equation}
However, in the star product $\star$ between components, we replace $\sigma$ or $\bar{\sigma}$ in the intermediate integration variables with $\tau$. \ie
\begin{equation}
(A_1\star \bar{B}_1)(\sigma_1,\sigma_2)\equiv\int d\tau_3 A_1(\sigma_1,\tau_3) d\tau_3 \bar{B}_1(\tau_3,\sigma_2)
\end{equation}
It is natural to consider chiral/chiral (or, anti-chiral/anti-chiral) supermatrices, too. They also follow the same multiplication rule in the supermatrix formulation. In general, the star product of supermatrices $A$ and $B$ is possible when the chirality of the second index of $A$ is the same as the chirality of the first index of $B$:
\begin{equation}
A_{u,v}\mstar^{v} B_{v,w}= C_{u,w}\qquad (u,v,w \in\{ \text{chiral}\;,\; \text{anti-chiral} \}) 
\end{equation}

Before discussing the $\mathcal{N}=2$ collective superfield theory, let us present useful formulae for the calculus of the bi-local superfield in $\mathcal{N}=2$ which generalize the formulae of Section~\ref{sec: n=2 calculus}. First, the functional derivative of the same fundamental superfield is given by
\begin{equation}
{\delta f(\sigma,\theta)\over \delta f(\sigma',\theta')}=(\theta'-\theta)\delta(\sigma'-\sigma)\quad,\quad {\delta \bar{f}(\bar{\sigma},\bar{\theta})\over \delta \bar{f}(\bar{\sigma}',\bar{\theta}')}=(\bar{\theta}'-\bar{\theta})\delta(\bar{\sigma}'-\bar{\sigma})
\end{equation}
We define the change of variables and chain rule for the fundamental superfield as follows.
\begin{align}
\delta f_\alpha(\sigma,\theta )=&\sum_\beta\int \delta f_\beta(\sigma',\theta') d\sigma' d\theta'  {\delta f_\alpha(\sigma, \theta)\over \delta f_\beta(\sigma',\theta')}\\
{\delta\over \delta f_\alpha(\sigma,\theta)}=&\sum_\beta\int {\delta  f_\beta(\sigma', \theta') \over \delta f_\alpha(\sigma,\theta)}d \sigma' d\theta' {\delta\over \delta f_\beta(\sigma', \theta')}\\
\delta \bar{f}_\alpha(\bar{\sigma},\bar{\theta} )=&\sum_\beta\int \delta \bar{f}_\beta(\bar{\sigma}',\bar{\theta}') d\bar{\sigma}' d\bar{\theta}'  {\delta \bar{f}_\alpha(\bar{\tau}, \bar{\theta})\over \delta \bar{f}_\beta(\bar{\tau}',\bar{\theta}')}\\
{\delta\over \delta \bar{f}_\alpha(\bar{\sigma},\bar{\theta})}=&\sum_\beta\int {\delta  \bar{f}_\beta(\bar{\sigma}', \bar{\theta}') \over \delta \bar{f}_\alpha(\bar{\sigma},\bar{\theta})}d \bar{\sigma}' d\bar{\theta}' {\delta\over \delta \bar{f}_\beta(\bar{\sigma}', \bar{\theta}')}
\end{align}
where $\alpha,\beta$ label some basis, and the summation runs over a complete basis. For bi-local superfields, we have the analogous formulae:
\begin{align}
{\delta F(\sigma_1,\theta_1;\bar{\sigma}_2,\bar{\theta}_2)\over \delta F(\sigma_3,\theta_3;\bar{\sigma}_4,\bar{\theta}_4)}\equiv& (\theta_3-\theta_1)(\bar{\theta}_4-\bar{\theta}_2)\delta(\sigma_3-\sigma_1)\delta(\bar{\sigma}_4-\bar{\sigma}_2)\\
{\delta \bar{F}(\bar{\sigma}_1,\bar{\theta}_1;\sigma_2,\theta_2)\over \delta \bar{F}(\bar{\sigma}_3,\bar{\theta}_3;\sigma_4,\theta_4)}\equiv& (\bar{\theta}_3-\bar{\theta}_1)(\theta_4-\theta_2)\delta(\bar{\sigma}_3-\bar{\sigma}_1)\delta(\sigma_4-\sigma_2)
\end{align}
\begin{align}
\delta F_\alpha(\sigma_1,\theta_1;\bar{\sigma}_2,\bar{\theta}_2)=& \sum_\beta\int  \delta F_\beta(\sigma_3,\theta_3;\bar{\sigma}_4,\bar{\theta}_4) d\bar{\sigma}_4d\bar{\theta}_4 d\sigma_3 d\theta_3 {\delta F_\alpha(\sigma_1,\theta_1;\bar{\sigma}_2,\bar{\theta}_2)\over \delta F_\beta(\sigma_3,\theta_3;\bar{\sigma}_4,\bar{\theta}_4)}\\
{\delta \over \delta F_\alpha(\sigma_1,\theta_1;\bar{\sigma}_2,\bar{\theta}_2)}= &\sum_\beta \int {\delta F_\beta(\sigma_3,\theta_3;\bar{\sigma}_4,\bar{\theta}_4) \over \delta F_\alpha(\sigma_1,\theta_1;\bar{\sigma}_2,\bar{\theta}_2)} d\bar{\sigma}_4 d\bar{\theta}_4  d\sigma_3 d\theta_3  {\delta \over \delta F_\beta(\sigma_3,\theta_3;\bar{\sigma}_4,\bar{\theta}_4) }\\
\delta \bar{F}_\alpha(\bar{\sigma}_1,\bar{\theta}_1;\sigma_2,\theta_2)=&\sum_\beta \int  \delta \bar{F}_\beta(\bar{\sigma}_3,\bar{\theta}_3;\sigma_4,\theta_4) d\sigma_4\theta_4 d\bar{\sigma}_3 d\bar{\theta}_3 {\delta \bar{F}_\alpha(\bar{\sigma}_1,\bar{\theta}_1;\sigma_2,\theta_2)\over \delta \bar{F}_\beta(\bar{\sigma}_3,\bar{\theta}_3;\sigma_4,\theta_4)}\\
{\delta \over \delta \bar{F}_\alpha(\bar{\sigma}_1,\bar{\theta}_1;\sigma_2,\theta_2)}= & \sum_\beta\int {\delta \bar{F}_\beta(\bar{\sigma}_3,\bar{\theta}_3;\sigma_4,\theta_4) \over \delta \bar{F}_\alpha(\bar{\sigma}_1,\bar{\theta}_1;\sigma_2,\theta_2)} d\sigma_4 d\theta_4  d\bar{\sigma}_3 d\bar{\theta}_3  {\delta \over \delta \bar{F}_\beta(\bar{\sigma}_3,\bar{\theta}_3;\sigma_4,\theta_4) }
\end{align}

\subsection{$\mathcal{N}=2$ Bi-local Collective Superfield Theory}
\label{sec: n=2 collective theory}

Consider Grassmannian odd chiral and anti-chiral superfields
\begin{equation}
\bar{\sD}\psi^i=0\quad,\quad \sD \bar{\psi}_i=0\hspace{1cm} (i=1,2,\cdots, N)
\end{equation}
In terms of component fields, we have
\begin{align}
\psi^i(\sigma,\theta)\equiv& \chi^i(\sigma)+\theta b^i(\tau)\hspace{1cm} (i=1,2,\cdots N)\\
\bar{\psi}_i(\bar{\sigma},\bar{\theta})\equiv& \bar{\chi}_i(\bar{\sigma})+\bar{\theta} \bar{b}_i(\tau)\hspace{1cm} (i=1,2,\cdots N)
\end{align}
where $\chi,\bar{\chi}$ are complex fermions while $b,\bar{b}$ are complex bosons. They transforms in the fundamental and anti-fundamental representation of $U(N)$, respectively:
\begin{equation}
\psi^i(\sigma,\theta) \;,\; \bar{\psi}_i(\bar{\sigma},\bar{\theta})  \quad \longrightarrow \quad {U^i}_j \psi^j(\sigma,\theta)\;,\; {\bar{U}_i \, }^j \bar{\psi}_j(\bar{\sigma},\bar{\theta})
\end{equation}
We define bi-local superfields and their conjugate:
\begin{align}
\Psi(\sigma_1,\theta_1;\bar{\sigma}_2,\bar{\theta}_2)\equiv& {1\over N} \psi^i(\sigma_1,\theta_1) \bar{\psi}_i(\bar{\sigma}_2,\bar{\theta}_2)\cr
\bar{\Psi}(\bar{\sigma}_1,\bar{\theta}_1;\sigma_2,\theta_2)\equiv& {1\over N} \bar{\psi}^i(\bar{\sigma}_1,\bar{\theta}_1) \psi_i(\sigma_2,\theta_2)\label{eq: n=2 anti-sym of superfield}
\end{align}
Note that $\Psi$ and $\bar{\Psi}$ are related by complex conjugation:
\begin{equation}
\overline{\left[\Psi(\sigma_1,\theta_1;\bar{\sigma}_2,\bar{\theta}_2)\right]}= -\bar{\Psi}(\sigma_1,\theta_1;\bar{\sigma}_2,\bar{\theta}_2)
\end{equation}
where this is not the complex conjugation of supermatrix but that of a superfield. As a supermatrix, it can be written as
\begin{equation}
\Psi(\sigma_1,\theta_1;\bar{\sigma}_2,\bar{\theta}_2)={1\over N}\begin{pmatrix}
b^i(\sigma_1,\theta_1)\bar{\chi}_i(\bar{\sigma}_2,\bar{\theta}_2) & -b^i(\sigma_1,\theta_1)\bar{b}_i(\bar{\sigma}_2,\bar{\theta}_2)\\
\chi^i(\sigma_1,\theta_1)\bar{\chi}_i(\bar{\sigma}_2,\bar{\theta}_2) & -\chi^i(\sigma_1,\theta_1)\bar{b}_i(\bar{\sigma}_2,\bar{\theta}_2)\\
\end{pmatrix}
\end{equation}
The complex conjugate relation of the bi-local superfields in~\eqref{eq: n=2 anti-sym of superfield} can be translated into the following relation in the supermatrix formulation.
\begin{equation}
\Jmatrix \Psi^\st \Jmatrix=\bar{\Psi}\label{eq: n=2 anti-sym of supermatrix}
\end{equation}
Hence, $\Psi$ and $\bar{\Psi}$ are not independent degrees of freedom, like a hermitian matrix. For the bi-local collective action, we need to evaluate a Jacobian coming from the non-trivial transformation of path integral measure. As in Section~\ref{sec:jacobian}, we will use the following identities for arbitrary functional $F[\Psi]$ of $\Psi$.
\begin{align}
&\int \mathcal{D}\psi \mathcal{D}\bar{\psi} {\delta\over \delta \psi^i(\sigma_1,\theta_1)}\left[ \psi^i(\sigma_2,\theta_2) F[\Psi] e^{-S}\right]=0\\
&\int \mathcal{D}\psi \int d\bar{\sigma}_3 d\bar{\theta}_3 {\delta\over \delta \Psi(\sigma_1,\theta_1;\bar{\sigma}_3,\bar{\theta}_3)}\left[ \Psi(\sigma_2,\theta_2;\bar{\sigma}_3,\bar{\theta}_3)\; \jacobian\; F[\Psi] e^{-S}\right]=0
\end{align}
and, is similar for $\bar{\Psi}$. In the same procedure as before, we can obtain functional differential equations for the Jacobian:
\begin{align}
N (\theta_1-\theta_2) \delta(\sigma_1-\sigma_2)=&\int  \Psi(\sigma_2,\theta_2;\bar{\sigma}_3,\bar{\theta}_3) d\bar{\sigma}_3 d\bar{\theta}_3 {\delta \log \jacobian \over \delta \Psi (\sigma_1,\theta_1;\bar{\sigma}_3,\bar{\theta}_3) } \\
N (\bar{\theta}_1-\bar{\theta}_2) \delta(\bar{\sigma}_1-\bar{\sigma}_2)=&\int  \bar{\Psi}(\bar{\sigma}_2,\bar{\theta}_2;\sigma_3,\theta_3) d\sigma_3 d\theta_3 {\delta \log \jacobian \over \delta \bar{\Psi} (\bar{\sigma}_1,\bar{\theta}_1;\sigma_3,\theta_3) }
\end{align}
As usual, this can be solved by
\begin{equation}
\log \jacobian=- {N\over 2} \str \log \Psi\bar{\mstar} \bar{\Psi}
\end{equation}
Note that the Jacobian $\jacobian$ should be a function of $ \Psi\bar{\mstar} \bar{\Psi}$ or $\bar{\Psi}\mstar \Psi $ because this is the only allowed combination, and they are related to
\begin{equation}
\log \jacobian=-{N\over 2}\str \log \Psi\bar{\mstar} \bar{\Psi}= -{N\over 2}\str \log (-\bar{\Psi}\mstar \Psi)
\end{equation}
Moreover, when analyzing the collective action later, one might be temped to treat $\Psi$ and $\bar{\Psi}$ as if they are independent variables. This seems to give the correct result, with certain prescriptions, as usual. However, rigorously speaking, they are not independent, and one should take this into account. For example, a functional derivative with respect to $\Psi$ will act on $\bar{\Psi}$ in the Jacobian. For this, it is helpful to use
\begin{equation}
\Jmatrix^\st (\Psi^\st)^\st \Jmatrix^\st=-\Jmatrix \Psi \Jmatrix
\end{equation}
in addition to the fact that supertrace is invariant under the supertranspose. Also, we do not have a shift in $N$ because the bi-local collective superfield does not have symmetry analogous to \eqref{eq:anti-sym of bilocal}. This was already seen in higher dimensional $U(N)$ vector models~\cite{deMelloKoch:1996mj,Das:2003vw,Jevicki:2014mfa}, and has been shown to be consistent for matching one-loop free energy of higher spin AdS/$U(N)$ vector model~\cite{Giombi:2013fka,Jevicki:2014mfa,Giombi:2014yra,Giombi:2014iua,Giombi:2016pvg}.

Now, to express the kinetic term, we will find the supermatrix representation of the superderivative.
\begin{equation}
\sD_1 A(\sigma_1,\theta_1,\bar{\sigma}_2,\bar{\theta}_2)
=\begin{pmatrix}
2\partial_{\tau_1} A_0(\tau_1;\bar{\sigma}_2)  & 2\partial_{\tau_1} A_2(\tau_1,\tau_2)  \\
A_1(\bar{\sigma}_1;\bar{\sigma}_2) & A_3(\bar{\sigma}_1;\tau_2)  \\
\end{pmatrix}\equiv \smD \mstar A
\end{equation}
Note the chiral superderivative is (Grassmannian odd) anti-chiral/chiral supermatrix:
\begin{equation}
\smD(\bar{\sigma}_1,\bar{\theta}_1;\sigma_2,\theta_2)\equiv 
\begin{pmatrix}
0 & 2 \partial_{\tau_1}\delta(\bar{\sigma}_1-\sigma_2) \\
\delta(\bar{\sigma}_1-\sigma_2) & 0\\
\end{pmatrix}
\end{equation}
Hence, the chiral superderivative can be multiplied to $\bar{\Psi}$ from the left by star product $\mstar$. In the same way, one can also define the anti-chiral superderivative as follows.
\begin{equation}
\bar{\smD}(\sigma_1,\theta_1;\bar{\sigma}_2,\bar{\theta}_2)\equiv
\begin{pmatrix}
0 & 2 \partial_{\tau_1}\delta(\sigma_1-\bar{\sigma}_2) \\
\delta(\sigma_1-\bar{\sigma}_2) & 0\\
\end{pmatrix}
\end{equation}
which satisfy
\begin{align}
(\bar{\smD}\bar{\mstar}\smD)(\sigma_1,\theta_1;\sigma_2,\theta_2)=2\partial_{\tau_1}\begin{pmatrix}
\delta(\sigma_1-\sigma_2) & 0\\
0 & \delta(\sigma_1-\sigma_2) \\
\end{pmatrix}= 2\partial_{\tau_1} \mathbb{I}(\sigma_1,\theta_1;\sigma_2,\theta_2)\\
(\smD\mstar\bar{\smD})(\bar{\sigma}_1,\bar{\theta}_1;\bar{\sigma}_2,\bar{\theta}_2)=2\partial_{\tau_1}\begin{pmatrix}
\delta(\bar{\sigma}_1-\bar{\sigma}_2) & 0\\
0 & \delta(\bar{\sigma}_1-\bar{\sigma}_2) \\
\end{pmatrix}= 2\partial_{\tau_1} \bar{\mathbb{I}}(\bar{\sigma}_1,\bar{\theta}_1;\bar{\sigma}_2,\bar{\theta}_2)
\end{align}
Then, in the supermatrix notation, the kinetic term can easily be written with the superderivative matrix as follows.
\begin{equation}
\str ( \smD\mstar \Psi)=\str ( \bar{\smD}\bar{\mstar} \bar{\Psi})=\int d\tau_1 \left[\left. 2 \partial_{\tau_1}\psi^i (\tau_1)\bar{\psi}_i(\tau_2)\right|_{\tau_2\rightarrow \tau_1 } + b^i(\tau_1) \bar{b}_i(\tau_1)\right]\label{eq:n=2 kinetic term}
\end{equation}
Therefore, like $\mathcal{N}=1$ case, the bi-local collective action for $\mathcal{N}=2$ SYK model is given by
\begin{align}
S_{col}=&-N \str \left(\smD\mstar  \Psi\right)+{N\over 2}\str \log (\Psi\bar{\mstar} \bar{\Psi})  - {JN\over 3}\int  d\tau_1d\theta_1 d\tau_2 d\bar{\theta}_2 [\Psi(\sigma_1,\theta_1;\bar{\sigma}_2,\bar{\theta}_2)]^3 \label{eq:n=2 col action superfield}\\
=&N \str \left[ - \smD\mstar  \Psi+ {1\over 2} \log (\Psi\bar{\mstar} \bar{\Psi}) -{J\over 3} \bar{\Psi}  \mstar [\Psi]^2 \right] \label{eq:n=2 col action supermatrix} 
\end{align}
The rest of calculation is parallel to $\mathcal{N}=1$ case except that the large $N$ classical solution need not to be anti-symmetric, which admits a one-parameter family of solutions depending on ``spectral asymmetry'' $\mathcal{E}$~\cite{Sachdev:2015efa,Davison:2016ngz}. Also, since the collective action as a supermatrix in~\eqref{eq:n=2 col action supermatrix} contains both $\Psi$ and $\bar{\Psi}$ which are not independent, one need additional care. Practically, it is useful to go back and forth between the supermatrix notation~\eqref{eq:n=2 col action supermatrix} and the superfield notation~\eqref{eq:n=2 col action superfield}. For example, the superfield notation is useful in varying the interaction term because one can easily change $\Psi$ into $\bar{\Psi}$. \ie
\begin{equation}
\int  d\tau_1d\theta_1 d\tau_2 d\bar{\theta}_2 [\Psi(\sigma_1,\theta_1;\bar{\sigma}_2,\bar{\theta}_2)]^3=\int  d\tau_1d\bar{\theta}_1 d\tau_2 d\theta_2 [\bar{\Psi}(\bar{\sigma}_1,\bar{\theta}_1;\sigma_2,\theta_2)]^3
\end{equation}
This is a trivial identity from the point of view of the superfield notation, which leads to an identity that can also be proven in the supermatrix notation:
\begin{equation}
\str\left[\bar{\Psi}  \mstar [\Psi]^2\right]=\str\left[\Psi  \mstar [\bar{\Psi}]^2\right]
\end{equation}
Varying the collective action with respect to $\Psi$ and multiplying $\Psi$ from the right, one can obtain the Schwinger-Dyson equation for the $\mathcal{N}=2$ SYK model~\cite{Fu:2016vas}:
\begin{equation}
-\smD\mstar \Psi +\mathbb{I}-[\Psi]^2\mstar \Psi =0
\end{equation}
One can also study $\mathcal{N}=2$ bi-local superconformal generators and its representation for the supermatrix formulation. Moreover, after finding the eigenfunctions for the Casimir operators, one can diagonalize the quadratic action to find all spectrum as in $\mathcal{N}=1$ SUSY SYK model. We leave them to future work.

\section{Conclusion}
\label{sec:conclusion}

In this work, we formulated the bi-local collective superfield theory for one-dimensional $\mathcal{N}=1,2$ SUSY vector models. We showed that this bi-local collective theory can be reformulated as supermatrix theory in the bi-local superspace. This drastically simplify the analysis of the $\mathcal{N}=1$ SUSY SYK model. We also studied the bi-local superconformal generators and its representation in the supermatrix formulation. Using them, we diagonalize the quadratic action of the $\mathcal{N}=1$ SUSY SYK model. We also developed the bi-local collective superfield theory for $\mathcal{N}=2$ SYK model, and also connected it to supermatrix formulation. The rich structures of the supermatrix formulation could provide deeper understanding on the SUSY SYK models.

In Section~\ref{sec:jacobian}, we easily obtain the shift in large $N$ by $-1$ which would be advantage of supersymmetry. Otherwise, one needs careful analysis of the differential equation for Jacobian. We showed that this shift in $N$ is not only important in matching free energy in the higher spin AdS/CFT but also in getting correct result in large $N$ expansion (See Appendix~\ref{app: correction}). Though we did not evaluate various observables by utilizing supersymmetry in this work, the simplicity of supermatrix formulation and the supersymmetry will enable us to calculate various observables exactly. We leave that to future work.

As mentioned in the introduction, this bi-local construction is not restricted to spacetime or superspace. The bi-local collective (super)field theory would shed light on the generalization of the SYK models like higher dimensional generalization by lattice. It is highly interesting to construct $\mathcal{N}=4$ bi-local superspace and its supermatrix formulation. Also, one might be able to generalize the bi-local superspace into higher-dimensional vector models in the context of higher spin AdS/CFT.

\acknowledgments

I thank Kimyeong Lee, Spenta Wadia, Antal Jevicki, R. Loganayagam, Prithvi Narayan, Victor Ivan Giraldo Rivera, and especially Robert de Mello Koch for extensive discussions. I would like to thank the Chennai Mathematical Institute for the hospitality and partial support during the early stages of the preparation of this work, within the program ``Student Talks on Trending Topics in Theory 2017''. I gratefully acknowledge support from International Centre for Theoretical Sciences (ICTS), Tata institute of fundamental research, Bengaluru. I would also like to acknowledge our debt to the people of India for their steady and generous support to research in the basic sciences.

\appendix

\section{${1\over N}$ Corrections in One-dimensional Free SUSY Vector Model}
\label{app: correction}

In this appendix, we show that the shift of $N$ by $-1$ indeed gives the correct one-point function of the bi-local collective superfield (or, invariant two-point function of the fundamental superfield) for a free theory. Consider a one-dimensional free vector model:
\begin{equation}
S_{\text{free}}=\int d\tau \left[{1\over 2} \chi^i \partial \chi^i -{1\over2} b^i b^i \right]\label{eq: action for free theory}
\end{equation}
Because it is a free theory, we expect the exact one-point function of the bi-local field will be
\begin{equation}
\langle \Psi(\tau_1,\theta_1;\tau_2,\theta_2) \rangle= \left\langle {1\over N} \psi^i(\tau_1,\theta_1)\psi^i(\tau_2,\theta_2)\right\rangle ={1\over 2}\left(\sgn(\tau_{12})-\theta_12\delta(\tau_{12})\theta_2 \right)
\end{equation}
The corresponding bi-local collective action for the free theory is given by
\begin{equation}
S_{col}=\str \left[ - {N\over 2}\smD\mstar  \Psi+ {N-1\over 2} \log \Psi \right]\label{eq: bi-local collective action for free theory}
\end{equation}
One can easily check that the large $N$ classical solution is the same as exact answer.
\begin{equation}
\Psi_{cl}(\tau_1,\theta_1;\tau_2,\theta_2)={1\over 2}\left(\sgn(\tau_{12})-\theta_12\delta(\tau_{12})\theta_2 \right)=\begin{pmatrix}
0 & \delta(\tau_{12})\\
{1\over 2} \sgn(\tau_{12}) & 0 \\
\end{pmatrix}\label{eq: classical solution for free}
\end{equation}
However, when we expand the bi-local superfield around the classical solution in large $N$
\begin{equation}
\Psi=\Psi_{cl}+ {1\over \sqrt{N} } \Phi
\end{equation}
the collective action~\eqref{eq: bi-local collective action for free theory} generates vertices which comes from
\begin{equation}
{N-1\over 2}\str \log \Psi
\end{equation} 
and, there should be no ${1\over N}$ correction to \eqref{eq: classical solution for free} from those vertices. At large $N$, the collective action can be expanded as
\begin{align}
S_{col}=&-{\sqrt{N}\over 2}\str (\smD \mstar \Phi) +{N-1\over 2} \sum_{m=1}^\infty {(-1)^{m+1}\over m N^{m\over 2}} \str \left[(\Psi_{cl}\mstar \Phi)^{\mstar m}\right]\cr
=&{\sqrt{N}\over 2} \str\left[ \Psi_{cl}^{-1} \mstar \Phi -\smD\mstar \Phi\right]-{1\over 4} \str ( \Psi_{cl}^{-1} \mstar \Phi \mstar  \Psi_{cl}^{-1} \mstar \Phi )\cr
& +{1\over 2\sqrt{N}}\str \left[- \Psi_{cl}^{-1} \mstar \Phi  +{1\over 3} \left( \Psi_{cl}^{-1} \mstar \Phi \right)^{\mstar 3}\right]+ \mathcal{O}(N^{-1})
\end{align}
First, one can easily calculate the inverse of the classical solution from \eqref{eq: classical solution for free}, and it turns out to be equal to the matrix superderivative in \eqref{def:bi-local superderivative}. 
\begin{equation}
\Psi_{cl}^{-1}=\smD
\end{equation}
In fact, this is the large $N$ Schwinger-Dyson equation for the free collective superfield theory. Then, from the quadratic action of order $\mathcal{O}(N^0)$, one can read off the two-point function of the bi-local fluctuation. Furthermore, one can easily show that
\begin{equation}
\langle ( \Psi_{cl}^{-1} \mstar \Phi \mstar  \Psi_{cl}^{-1} \mstar \Phi )(\tau_1,\tau_2)\rangle =\begin{pmatrix}
\delta(\tau_{12}) & 0\\
0 & \delta(\tau_{12})\\
\end{pmatrix}\label{eq: app identity 1}
\end{equation}
Now, the leading correction to the one-point function of the bi-local collective superfield is given by
\begin{equation}
{1\over 2N}\left\langle \Phi(\tau_1,\theta_1;\tau_2,\theta_2) \str \left[- \Psi_{cl}^{-1} \mstar \Phi  +{1\over 3} \left( \Psi_{cl}^{-1} \mstar \Phi \right)^{\mstar 3}\right] \right\rangle
\end{equation}
Using a property of the supertrace and \eqref{eq: app identity 1}, one can easily see that this correction vanishes. If it were not for the shift in $N$ by $(-1)$, this correction would not vanish, and therefore would not give the exact one-point function which one can expect in free theory. Though this shift does not have any influence in the main text of this paper, it would be important in evaluating ${1\over N}$ corrections to correlation functions or the free energy.

\section{Casimir Eigenfunctions}
\label{app: casimir eigenfunction}

In this appendix, we present the (bosonic and fermionic) eigenfunctions of the superconformal Casimir operators discussed in Section~\ref{sec: eigenfunction of casimir}.

\subsection{Bosonic Eigenfunctions}
\label{app: bosonic eigenfunction}

\begin{itemize}
	\item {\bf Eigenvalue of Casimir:} $\boldsymbol{\nu\left(\nu-{1\over 2}\right)}$
	
	\begin{align}
      \Gamma^1_{\nu w}=&e^{-iwt} z^{1\over 6} J_\nu (wz) 
      \begin{pmatrix} 
      0 & -{\nu-{1\over 6}\over 2 z }  \\
      1   &0 \\
      \end{pmatrix} \label{eq:eigenfunction1}\\
       \Gamma^2_{\nu w}=&e^{-iwt} z^{1\over 6}J_{-\nu}(wz) \begin{pmatrix}
    0 &  -{\nu - {1\over 6}\over 2 z}  \\
    1 & 0 \\
    \end{pmatrix}\quad\mbox{or}\quad \Gamma^2_{\nu w}=e^{-iwt} z^{1\over 6} Y_\nu (wz)\begin{pmatrix}
        0 & -{ \nu-{1\over 6}\over 2z} \\
        1 & 0\\
        \end{pmatrix}\qquad 
     \label{eq:eigenfunction1--2}\\
     \Gamma^3_{\nu w}=&{i\over 2} we^{-iw t} z^{1\over 6} \left[J_\nu(wz) \mathbb{1} + i J_{\nu-1}(wz) \boldsymbol{\sigma}_3 \right]\\
         \Gamma^4_{\nu w}= & {i\over 2} we^{-iw t} z^{1\over 6} \left[Y_\nu(wz) \mathbb{1} + i Y_{\nu-1}(wz) \boldsymbol{\sigma}_3 \right]\cr
    \mbox{or} \qquad& \Gamma^4_{\nu w}= {i\over 2} we^{-iw t} z^{1\over 6} \left[J_{-\nu}(wz) \mathbb{1} -  i J_{-\nu+1}(wz) \boldsymbol{\sigma}_3 \right]
    \end{align}

    \item {\bf Eigenvalue of Casimir:} $\boldsymbol{\nu\left(\nu+{1\over 2}\right)}$
    
    \begin{align}
        \Gamma^5_{\nu w}=&e^{-iwt} z^{1\over 6} J_{\nu} (wz)
        \begin{pmatrix}
        0 & {\nu+{1\over 6} \over 2 z}\\
        1 &  0 \\
         \end{pmatrix} \label{eq:eigenfunction2}\\
          \Gamma^6_{\nu w}=&e^{-iwt} z^{1\over 6}J_{-\nu} (wz)  \begin{pmatrix}
        0 & {\nu+{1\over 6}\over 2z} \\
        1 & 0 \\
        \end{pmatrix}\quad\mbox{or}\quad \Gamma^6_{\nu w}=e^{-iwt} z^{1\over 6}Y_{\nu} (wz)  \begin{pmatrix}
        0 & {\nu+{1\over 6}\over 2z} \\
        1 & 0 \\
        \end{pmatrix}\label{eq:eigenfunction2--1}\\
         \Gamma^7_{\nu w}=&{i\over 2} we^{-iw t} z^{1\over 6} \left[J_\nu(wz) \mathbb{1} - i J_{\nu+1}(wz) \boldsymbol{\sigma}_3 \right]\\
          \Gamma^8_{\nu w}= & {i\over 2} we^{-iw t} z^{1\over 6} \left[Y_\nu(wz) \mathbb{1} - i Y_{\nu+1}(wz) \boldsymbol{\sigma}_3 \right]\cr
     \mbox{or}\qquad&\Gamma^8_{\nu w}= {i\over 2} we^{-iw t} z^{1\over 6} \left[J_{-\nu}(wz) \mathbb{1} +  i J_{-\nu-1}(wz) \boldsymbol{\sigma}_3 \right] 
    \end{align}

\item {\bf Action of Supercharge:}

    \begin{align}
    \dQ \Gamma^1_{\nu w}=&(iw) {i\over 2}  e^{-iwt} z^{1\over 6} \left[J_{\nu-1}(wz)\mathbb{1} - i J_\nu(wz)\boldsymbol{\sigma}_3\right]\\
    \dQ \Gamma^3_{\nu w}=&(iw) {i\over 2}  e^{-iwt} z^{1\over 6} J_{\nu+1}(wz)\begin{pmatrix}
    0 & {\nu-{1\over 6}\over 2z} \\
    1  &  0\\
    \end{pmatrix}\\
       \dQ \Gamma^5_{\nu w}=&(iw) {i\over 2}  e^{-iwt} z^{1\over 6} \left[-J_{\nu+1}(wz)\mathbb{1} - i J_\nu(wz)\boldsymbol{\sigma}_3\right]\\
        \dQ \Gamma^7_{\nu w}=&(iw) {i\over 2}  e^{-iwt} z^{1\over 6} J_{\nu}(wz)\begin{pmatrix}
    0 & -{\nu+{1\over 6}\over 2z} \\
    1  &  0\\
    \end{pmatrix}
    \end{align}

\end{itemize}

\subsection{Fermionic Eigenfunctions}
\label{app: fermionic eigenfunction}

\begin{itemize}

	\item {\bf Eigenvalue of Casimir:} $\boldsymbol{\nu\left(\nu-{1\over 2}\right)}$
	
	\begin{align}
      \Omega^1_{\nu w}=&e^{-iwt} z^{1\over 6} J_\nu (wz) 
      \begin{pmatrix} 
      0 & {\nu-{1\over 6}\over 2 z }  \\
      1   &0 \\
      \end{pmatrix} \label{eq:eigenfunction3--1}\\
       \Omega^2_{\nu w}=&e^{-iwt} z^{1\over 6}J_{-\nu}(wz) \begin{pmatrix}
    0 &  {\nu - {1\over 6}\over 2 z}  \\
    1 & 0 \\
    \end{pmatrix}\quad\mbox{or}\quad \Omega^2_{\nu w}=e^{-iwt} z^{1\over 6} Y_\nu (wz)\begin{pmatrix}
        0 & { \nu-{1\over 6}\over 2z} \\
        1 & 0\\
        \end{pmatrix}\qquad 
     \label{eq:eigenfunction3--2}\\
     \Omega^3_{\nu w}=&{i\over 2} we^{-iw t} z^{1\over 6} \left[J_\nu(wz) \mathbb{1} + i J_{\nu-1}(wz) \boldsymbol{\sigma}_3 \right]\\
         \Omega^4_{\nu w}= & {i\over 2} we^{-iw t} z^{1\over 6} \left[Y_\nu(wz) \mathbb{1} + i Y_{\nu-1}(wz) \boldsymbol{\sigma}_3 \right]\cr
    \mbox{or} \qquad& \Omega^4_{\nu w}= {i\over 2} we^{-iw t} z^{1\over 6} \left[J_{-\nu}(wz) \mathbb{1} -  i J_{-\nu+1}(wz) \boldsymbol{\sigma}_3 \right]
    \end{align}

	\item {\bf Eigenvalue of Casimir:} $\boldsymbol{\nu\left(\nu+{1\over 2}\right)}$
	
  \begin{align}
        \Omega^5_{\nu w}=&e^{-iwt} z^{1\over 6} J_{\nu} (wz)
        \begin{pmatrix}
        0 & -{\nu+{1\over 6} \over 2 z}\\
        1 &  0 \\
         \end{pmatrix} \label{eq:eigenfunction4}\\
          \Omega^6_{\nu w}=&e^{-iwt} z^{1\over 6}J_{-\nu} (wz)  \begin{pmatrix}
        0 & -{\nu+{1\over 6}\over 2z} \\
        1 & 0 \\
        \end{pmatrix}\quad\mbox{or}\quad \Gamma^6_{\nu w}=e^{-iwt} z^{1\over 6}Y_{\nu} (wz)  \begin{pmatrix}
        0 & -{\nu+{1\over 6}\over 2z} \\
        1 & 0 \\
        \end{pmatrix}\label{eq:eigenfunction4--2}\\
         \Omega^7_{\nu w}=&{i\over 2} we^{-iw t} z^{1\over 6} \left[J_\nu(wz) \mathbb{1} - i J_{\nu+1}(wz) \boldsymbol{\sigma}_3 \right]\\
          \Omega^8_{\nu w}= & {i\over 2} we^{-iw t} z^{1\over 6} \left[Y_\nu(wz) \mathbb{1} - i Y_{\nu+1}(wz) \boldsymbol{\sigma}_3 \right]\cr
     \mbox{or}\qquad&\Omega^8_{\nu w}= {i\over 2} we^{-iw t} z^{1\over 6} \left[J_{-\nu}(wz) \mathbb{1} +  i J_{-\nu-1}(wz) \boldsymbol{\sigma}_3 \right] 
    \end{align}

\item {\bf Action of Supercharge:}

    \begin{align}
     \dQ \Omega^1_{\nu w}=&(iw) {i\over 2}  e^{-iwt} z^{1\over 6} \left[-iJ_{\nu}(wz)\mathbb{1} + J_{\nu-1}(wz)\boldsymbol{\sigma}_3\right]\\
    \dQ \Omega^3_{\nu w}=&(iw) {i\over 2}  e^{-iwt} z^{1\over 6} J_{\nu}(wz)\begin{pmatrix}
    0 & {\nu-{1\over 6}\over 2z} \\
    1  &  0\\
    \end{pmatrix}\\
     \dQ \Omega^5_{\nu w}=&(iw) {i\over 2}  e^{-iwt} z^{1\over 6} \left[-iJ_{\nu}(wz)\mathbb{1} -  J_{\nu+1}(wz)\boldsymbol{\sigma}_3\right]\\
        \dQ \Omega^7_{\nu w}=&(iw) {i\over 2}  e^{-iwt} z^{1\over 6} J_{\nu}(wz)\begin{pmatrix}
    0 & -{\nu+{1\over 6}\over 2z} \\
    1  &  0\\
    \end{pmatrix}
    \end{align}

\end{itemize}

\section{Direct Diagonalization}
\label{app: diagonalization}

In this Appendix, we will diagonalize the quadratic action following~\cite{Polchinski:2016xgd,Jevicki:2016bwu}. In~\ref{sec: eigenfunctions of the quadratic action}, we already showed that the second term in the quadratic action~\eqref{eq:quadratic action supermatrix} corresponds to the inner product of two eigenfunctions. Hence, we will focus on the first term of the quadratic action. For each $u^a_{\nu w}$ ($a=1,2$), we will find $\tilde{u}^a_{\nu w}$ such that 
\begin{equation}
\Psi_{cl}\mstar \tilde{u}_{\nu w} \mstar \Psi_{cl}\mstar= g(\nu) u_{\nu w}\label{eq: app equation}
\end{equation}
where we will use the known functions $g(\nu)$'s in \cite{Fu:2016vas}. (See \eqref{eq:spectrum eq1} and \eqref{eq:spectrum eq2}.) Because of the symmetry of $\tilde{u}_{\nu w}$ in~\eqref{eq: u tilde sym}, we have the following ansatz.
\begin{equation}
    \tilde{u}_{\nu w}(\tau_1,\tau_2)\sim \begin{pmatrix}
    0 & \mu f^a_{7\over 6}(\tau_{12}) \\
    f^s_{1\over 6}(\tau_{12}) & 0\\
    \end{pmatrix} 
\end{equation}
One component of the LHS in \eqref{eq: app equation} is 
\begin{align}
    &\int {d\tau_3 d\tau_4\over |{1\over 2}(\tau_3-\tau_4)|^{7\over 6} } \; {\sgn(\tau_{13})\sgn(\tau_{42})  Z_{\nu}(\left|{w\over 2}(\tau_3-\tau_4)\right|) \sgn(\tau_3-\tau_4)\over |\tau_1-\tau_3|^{1\over 3}|\tau_4-\tau_2|^{1\over 3}}\cr
    =&-2 e^{-iwt_0}\int dtdz  {|z-z_0|^{1\over 3} Z_{\nu}(|wz|) \sgn(z) \over  |z|^{7\over 6}  }  {e^{-iw|z-z_0| t}\sgn(t+1)\sgn(t-1) \over |t^2-1|^{1\over 3}}\cr
    =&-4  e^{-iwt_0}\int dz  {|z-z_0|^{1\over 3} Z_{\nu}(|wz|)\sgn(z) \over  |z|^{7\over 6}  } \cr
    &\hspace{2cm}\times\left[\int_1^\infty dt {\cos w|z-z_0|t\over |t^2-1|^{1\over 3}}-\int_0^1 dt {\cos w|z-z_0|t\over |1-t^2|^{1\over 3}}\right]\cr
    =&2\sqrt{\pi} \left({2\over w}\right)^{1\over 6}\Gamma({2\over 3})e^{-iwt_0}\int dz  {|z-z_0|^{1\over 6} Z_{\nu}(|wz|) \sgn(z)\over  |z|^{7\over 6}  }\cr
    &\hspace{2cm}\times \left[J_{1\over 6}(|w(z-z_0)|)+Y_{-{1\over 6}}(|w(z-z_0)|)\right]\label{eq:component 1}
\end{align}
up to a trivial factor. Here, we defined
\begin{align}
t\equiv{1\over 2}(\tau_3+\tau_4)\quad&,\quad z\equiv {1\over 2} (\tau_3-\tau_4)\\
t_0\equiv{1\over 2}(\tau_1+\tau_2)\quad&,\quad z_0\equiv{1\over 2} (\tau_1-\tau_2)
\end{align}
In the last line, we used eq.~(3.771) in~\cite{integral}:
\begin{align}
    \int_0^1 dx {\cos ax\over (x^2-1)^b }=&{\sqrt{\pi}\over 2}\left({a\over 2}\right)^{b-{1\over 2}}\Gamma(1-b)J_{{1\over 2}-b}(a)\quad(a>0,\; \Re b<1)\\
    \int_1^\infty dx {\cos ax\over (1-x^2)^b }=&-{\sqrt{\pi}\over 2}\left({a\over 2}\right)^{b-{1\over 2}}\Gamma(1-b)Y_{b-{1\over 2}}(a)\quad(a>0,\; \Re b>0)
\end{align}
In the same way, we found that the other component becomes
\begin{align}
    &\int {d\tau_3 d\tau_4\over |{1\over 2}(\tau_3-\tau_4)|^{1\over 6} } \; {  Z_{\nu}(\left|{w\over 2}(\tau_3-\tau_4)\right|) \over |\tau_1-\tau_3|^{4\over 3}|\tau_4-\tau_2|^{4\over 3}}=2 e^{-iwt_0}\int {dtdz\over |z|^{1\over 6}} {e^{-iwt} Z_{\nu}(|wz|)\over |t^2-(z-z_0)^2|^{4\over 3} }\cr
    =&{2\over 9} A J^{5\over 6}c^2 e^{-iwt_0}\int dtdz  { Z_{\nu}(|wz|)  \over  |z|^{1\over 6} |z-z_0|^{5\over 3} }  {e^{-iw|z-z_0| t} \over |t^2-1|^{4\over 3}}\cr
    =&4 e^{-iwt_0}\int dz  { Z_{\nu}(|wz|) \over  |z|^{1\over 6}|z-z_0|^{5\over 3}  } \cr
    &\hspace{2cm}\times\left[\int_1^\infty dt {\cos w|z-z_0|t\over |t^2-1|^{4\over 3}}+\int_0^1 dt {\cos w|z-z_0|t\over |1-t^2|^{4\over 3}}\right]\cr
    =&2\sqrt{\pi}  \left({w\over 2}\right)^{5\over 6}\Gamma(-{1\over 3})e^{-iwt_0}\int dz  { Z_{\nu}(|wz|) \over  |z|^{1\over 6} |z-z_0|^{5\over 6} }\cr
    &\hspace{2cm}\times \left[J_{-{5\over 6}}(|w(z-z_0)|)-Y_{{5\over 6}}(|w(z-z_0)|)\right]\label{eq:component 2}
\end{align}
up to trivial factors.

Now, we will use Fourier transformation of each Bessel function with appropriate factors. That is, in the LHS of \eqref{eq: app equation}, we will consider the Fourier transformations of the following six functions.
\begin{align}
|z-z_0|^{1\over 6} J_{1\over 6}(|w(z-z_0)|)\;,\; |z-z_0|^{1\over 6} Y_{-{1\over 6}}(|w(z-z_0)|)\;, \; |z|^{-{7\over 6}} Z_{\nu}(|wz|)\\
|z-z_0|^{-{5\over 6}} J_{-{5\over 6}}(|w(z-z_0)|)\;,\; |z-z_0|^{-{5\over 6}} Y_{{5\over 6}}(|w(z-z_0)|)\;, \; |z|^{-{1\over 6}} Z_{\nu}(|wz|)
\end{align}
while on the RHS we need the Fourier transformation of the following function.
\begin{equation}
|z_0|^{1\over 6}Z_\nu (|wz_0|)\
\end{equation}
The Fourier transformation of these functions can be performed by using the following integrals~(\eg See eq.~(6.699) in \cite{integral}).
\begin{align}
    &\Romannum{1}\;:\;\int dx\; x^\nu e^{ipx} J_\nu(|x|)=2\int dx\;x^\nu \cos px  J_\nu (x)\cr
    &\hspace{2cm}={2^{1+\nu} \Gamma({1\over 2}+\nu )\over \sqrt{\pi} |p^2-1|^{\nu+{1\over 2}} }\left[\theta(1-|p|)-\sin\pi\nu \theta(|p|-1)\right] \\
    &\Romannum{2}\;:\;\int dx\; x^\nu e^{ipx} J_{-\nu}(|x|)=2\int dx\;x^\nu \cos px  J_{-\nu} (x)\cr
    &\hspace{2cm}={2^{1+\nu}\sqrt{\pi} \over \Gamma({1\over 2}-\nu )}{}_2F_1({1\over 2},{1\over 2}+\nu,{1\over 2};p^2) \theta(1-|p|)
\end{align}
\begin{align}
    \Romannum{3}\;:\;&\int dx\; x^\nu e^{ipx} Y_{\nu}(|x|)=2\int dx\;x^\nu \cos px  {\cos \pi \nu J_\nu(x) - J_{-\nu} (x)\over \sin \pi \nu}\cr
    =&{2^\nu \Gamma({1\over 2}+\nu )\over \sqrt{\pi} |p^2-1|^{\nu+{1\over 2}} }\left[\theta(1-|p|)-\sin\pi\nu \theta(|p|-1)\right] \cr
    &-{2^{1-\nu}\sqrt{\pi}|p|^{2\nu+1} \over \sin \pi \nu\Gamma({1\over 2}+\nu ) |p^2-1|^{\nu+{1\over 2}} } \theta(|p|-1)\\
    \Romannum{4}\;:\;&\int dx \; |x|^\mu e^{ipx} J_\nu(|x|)= 2\int_0^\infty dx\;  x^\mu \cos px J_\nu (x)\cr
    =&{2^{1-\nu} \Gamma\left(1+\mu+\nu\right)\cos\left[{\pi\over 2}(1+\mu+\nu)\right]\over \Gamma\left(\nu+1\right)|p|^{1+\mu+\nu}}F\left({1+\mu+\nu\over 2},{2+\mu+\nu\over 2},\nu+1;{1\over p^2}\right)\theta(|p|-1)\cr
    &+ {2^{1+\mu} \Gamma\left({1+\mu+\nu\over 2}\right)\over \Gamma\left({\nu-\mu+1\over 2}\right)}F\left({1+\mu+\nu\over 2},{1+\mu-\nu\over 2},{1\over 2}; p^2\right)\theta(1-|p|) 
\end{align}
\begin{align}
    \Romannum{5}\;:\;&\int dx \; |x|^\mu e^{ipx} J_{-\nu}(|x|)= 2\int_0^\infty dx\;  x^\mu \cos px J_{-\nu} (x)\cr
    =&{2^{1+\nu} \Gamma\left(1+\mu-\nu\right)\cos\left[{\pi\over 2}(1+\mu-\nu)\right]\over \Gamma\left(-\nu+1\right)|p|^{1+\mu-\nu}}F\left({1+\mu-\nu\over 2},{2+\mu-\nu\over 2},-\nu+1;{1\over p^2}\right)\theta(|p|-1)\cr
    &+ {2^{1+\mu} \Gamma\left({1+\mu-\nu\over 2}\right)\over \Gamma\left({-\nu-\mu+1\over 2}\right)}F\left({1+\mu-\nu\over 2},{1+\mu+\nu\over 2},{1\over 2};p^2\right)\theta(1-|p|)\\
    \Romannum{6}\;:\;&\int dx \; |x|^\mu \sgn(z) e^{ipx} J_\nu(|x|)= 2i\int_0^\infty dx\;  x^\mu \sin px J_\nu (x)\cr
    =& i2^{1-\nu} {\Gamma(\nu+\mu+1) \sin\left[{\pi\over 2}(1+\mu+\nu)\right]\over \Gamma(\nu+1) |p|^{\nu+\mu+1}} F\left({2+\mu+\nu\over 2},{1+\mu+\nu\over 2},\nu+1;{1\over p^2}\right)\theta(|p|-1)\cr
    &+ i2^{2+\mu}\sgn(p) |p| {\Gamma\left({2+\mu+\nu\over 2}\right)\over \Gamma\left({\nu-\mu\over 2}\right)} F\left({2+\mu+\nu\over 2},{2+\mu-\nu\over 2},{3\over 2};p^2\right)\theta(1-|p|)
\end{align}
Substituting these Fourier modes into \eqref{eq:component 1} and \eqref{eq:component 2}, one can perform the integration with respect to $z$. The $e^{-iw t_0}$ factor can be easily obtained. By comparing the rest of the components on the both sides of~\eqref{eq: app equation}, we found that 
\begin{align}
\mu=-{1\over 2}\left(\nu+{1\over 6}\right)\qquad &\mbox{for}\;\; u^1_{\nu w}\\
\mu={1\over 2}\left(\nu-{1\over 6}\right)\qquad &\mbox{for}\;\; u^2_{\nu w}
\end{align}
and, thus we also confirmed our claim in~\eqref{eq: tilde solution}. Using there $u_{\nu w}$'s, we obtain the eigenvalues of the kernel by evaluating the inner product. We find that
\begin{align}
    -{ J\nu \over 2^{2\over 3} 3^{1\over 6} \pi^{1\over 3} }\widetilde{N}^-_\nu\left({1\over g_1(\nu)}-1\right)\qquad &\mbox{for}\;\; u^1_{\nu w}\\
      { J\nu \over 2^{2\over 3} 3^{1\over 6} \pi^{1\over 3} }\widetilde{N}^+_\nu\left({1\over g_2(\nu)}-1\right)\qquad &\mbox{for}\;\; u^2_{\nu w}
\end{align}

Now, we will confirm a part of diagonalization of the quadratic action (\ie the second term in \eqref{eq:quadratic action supermatrix}) by $ \dQ \mB\mstar u^a_{\nu w}$ $(a=3,4)$. Explicitly, we obtain
\begin{equation}
    \dQ \mB\mstar u^3_{\nu w} 
    ={|w|\over 2} {1\over \sqrt{8\pi}} |Jz|^{1\over 6}\begin{cases}  \;e^{-i|w|t } \left[ Z^+_{\nu-1}(|wz|)\sgn(z) \boldsymbol{\sigma_3} - iZ^-_\nu(|wz|)\mB\mstar  \right] &\quad (w>0)\\
    \;e^{i|w|t } \left[  Z^+_{\nu-1}(|wz|)\sgn(z)\mB\mstar\boldsymbol{\sigma_3} + iZ^-_\nu(|wz|)\mB\mstar \right] &\quad (w<0)
    \end{cases}
\end{equation}
where $\boldsymbol{\sigma_3}$ is a Pauli-like supermatrix (\ie $\boldsymbol{\sigma_3}=\begin{pmatrix}
1 & 0 \\
0 & -1\\
\end{pmatrix}$) and
\begin{align}
    \nu=&{3\over 2}+2n\hspace{1cm} (n=0,1,2,\cdots)\\
    \nu=&ir\hspace{1cm} (r\in \mathbb{R})
\end{align}
In component, we have
\begin{align}
    v^{11}_{\nu w}=&\oscB{|w|\over 2} {1\over \sqrt{8\pi}} |Jz|^{1\over 6}\begin{cases}  \;e^{-i|w|t } \left[ Z^+_{\nu-1}(|wz|)\sgn(z)-iZ^-_\nu(|wz|) \right] &\quad (w>0)\\
    \;e^{i|w|t } \left[ Z^+_{\nu-1}(|wz|)\sgn(z) + iZ^-_\nu(|wz|)  \right] &\quad (w<0)
    \end{cases} \\
    v^{12}_{\nu w}=&\oscB{|w|\over 2} {1\over \sqrt{8\pi}} |Jz|^{1\over 6}\begin{cases}  \;e^{-i|w|t } \left[ -Z^+_{\nu-1}(|wz|)\sgn(z)-iZ^-_\nu(|wz|)  \right] &\quad (w>0)\\
    \;e^{i|w|t } \left[ - Z^+_{\nu-1}(|wz|)\sgn(z) + iZ^-_\nu(|wz|)  \right] &\quad (w<0)
    \end{cases}
\end{align}
We will evaluate
\begin{equation}
2J^{4\over 3}c \int d\tau_1d\tau_2\; f^a_{1/3}\eta_1(\tau_1,\tau_2)\eta_2(\tau_1,\tau_2)\label{eq:fermi second term}
\end{equation}
where we expand the $\eta$'s in terms of $v^{11}$ and $v^{12}$. \ie
\begin{align}
    \eta_1=&\sum_{w\geqq 0}\sum_{\substack{\nu=ir\\ r \geqq 0}} \oscB^1_{\nu w} v^{11}_{\nu w}+\sum_{w\geqq 0}\sum_{\substack{\nu=2n+{3\over 2} \\ n=0,1,\cdots}}\oscB^1_{\nu w} v^{11}_{\nu w} +\mbox{c.c.}\\
    \eta_2=&\sum_{w\geqq 0}\sum_{\substack{\nu=ir\\ r \geqq 0 }} \oscB^1_{\nu w} v^{12}_{\nu w}+\sum_{w\geqq 0}\sum_{\substack{\nu=2n+{3\over 2} \\ n=0,1,\cdots}}\oscB^1_{\nu w} v^{12}_{\nu w} +\mbox{c.c.}
\end{align}
In order to evaluate these integrals, we need an identity
\begin{equation}
    Z^+_{\nu-1}=J_{\nu-1}+\xi_{-\nu+1}J_{-\nu+1}
    =\partial_z Z^-_\nu +{\nu\over z}Z^-_\nu\label{app: eigenfunc identity}
\end{equation}
where we used
\begin{equation}
    \xi_{-\nu+1}=-\xi_\nu \ .
\end{equation}
The identity \eqref{app: eigenfunc identity} enables us to evaluate the following integral.
\begin{align}
    \int_0^\infty dz \;(Z^+_{\nu-1}(z)Z^-_{\mu}(z)+Z^-_{\nu}(z)Z^+_{\mu-1}(z))=&\left. Z^-_\nu Z^-_\mu \right|_0^\infty +(\nu+\mu)\int_0^\infty {dz\over z} Z^-_\nu(z)Z^-_{\mu}(z)\cr
    =&2 \nu\widetilde{N}^-_\nu \delta(\nu-\mu)
\end{align}
Then, we find that \eqref{eq:fermi second term} is
\begin{align}
    &2J^{4\over 3}c \int d\tau_1d\tau_2\; f^a_{1/3}\eta_1(\tau_1,\tau_2)\eta_2(\tau_1,\tau_2)\cr
    &\hspace{2cm}={2J \over 2^{2\over 3} 3^{1\over 6}\pi^{1\over 6} } \left[\sum_{\substack{\nu=ir\\r\geqq 0}}+ \sum_{\substack{\nu=2n+{3\over 2} \\ n=0,1,\cdots}} \right]\sum_{w\geqq 0} B^1_{\nu w}B^1_{\nu, -w} (-iw) \nu \widetilde{N}^-_\nu
\end{align}
For the other modes, one can repeat the same evaluation. $\dQ u^4_{\nu w}$ is given by
\begin{equation}
    \dQ u^4_{\nu w} 
    ={|w|\over 2} {1\over \sqrt{8\pi}} |Jz|^{1\over 6}\begin{cases}  \;e^{-i|w|t } \left[ Z^-_{\nu+1}(|wz|)\sgn(z)\boldsymbol{\sigma_3} +iZ^+_\nu(|wz|)\mathbb{1} \right] &\quad (w>0)\\
    \;e^{i|w|t } \left[  Z^-_{\nu+1}(|wz|)\sgn(z)\boldsymbol{\sigma_3} -iZ^+_\nu(|wz|)\mathbb{1} \right] &\quad (w<0)
    \end{cases}
\end{equation}
where
\begin{align}
    \nu=&{1\over 2}+2n\hspace{1cm} (n=0,1,2,\cdots)\\
    \nu=&ir\hspace{1cm} (r\in \mathbb{R})
\end{align}
In components, we have
\begin{align}
    v^{21}_{\nu w}=&{|w|\over 2} {1\over \sqrt{8\pi}} |Jz|^{1\over 6}\begin{cases}  \;e^{-i|w|t } \left[ Z^+_{\nu+1}(|wz|)\sgn(z)+iZ^-_\nu(|wz|) \right] &\quad (w>0)\\
    \;e^{i|w|t } \left[ Z^+_{\nu+1}(|wz|)\sgn(z) - iZ^-_\nu(|wz|)  \right] &\quad (w<0)
    \end{cases} \\
    v^{22}_{\nu w}=&{|w|\over 2} {1\over \sqrt{8\pi}} |Jz|^{1\over 6}\begin{cases}  \;e^{-i|w|t } \left[ -Z^+_{\nu+1}(|wz|)\sgn(z)+ iZ^-_\nu(|wz|)  \right] &\quad (w>0)\\
    \;e^{i|w|t } \left[ - Z^+_{\nu+1}(|wz|)\sgn(z) - iZ^-_\nu(|wz|)  \right] &\quad (w<0)
    \end{cases}
\end{align}

\section{$\mathcal{N}=1$ SUSY SYK model: General $q$ }
\label{app: general q}

In this appendix, we discuss the eigenvectors of the $\mathcal{N}=1$ SUSY SYK model for the general $q$ case. Since the idea is the same as the $q=3$ case, we present only important results. For the general $q$ case, since the fundamental superfield has dimension ${1\over 2q}$, the appropriate $\mathcal{N}=1$ superconformal generators are given by
\begin{align}
\dP_a=&\partial_{\tau_a}\\
\dK_a=&\tau_a^2\partial_{\tau_a} + 2\Delta_a \tau_a +\tau_a \theta_a\partial_{\theta_a}\\
\dD_a=&\tau_a\partial_{\tau_a}  +{1\over 2} \theta_a \partial_{\theta_a} +\Delta_a\\
\dQ_a=&\partial_{\theta_a} -\theta_a\partial_{\tau_a}\\
\dS_a=&\tau_a\partial_{\theta_a} - \tau_a \theta_a \partial_{\tau_a} - 2\Delta_a \theta_a
\end{align}
where $a=1,2$ and $ \Delta_a\equiv{1\over 2q}$ $(a=1,2)$. The bi-local superconformal generators are defined by
\begin{equation}
    \mathcal{L}=\mathcal{L}_1+\mathcal{L}_2\hspace{1cm} (\; L\in \{P,K,D,\dQ,S\}\;)
\end{equation}
and the associated Casimir is
\begin{align}
\mathcal{C}=&\dD^2-{1\over2}(\dP\dK+\dK\dP)+{1\over 4} (\dS\dQ-\dQ \dS)=\dD^2-{1\over2} \dD-\dK\dP+{1\over 2} \dS\dQ
\end{align}
Via the bi-local map in~\eqref{eq:bi-local map} and \eqref{eq:bi-local map2}, the superconformal generators are represented as
\begin{align}
    \dP=&\partial_t\\
    \dK=&(t^2+z^2)\partial_t+2tz\partial_z+t(\zeta_0\partial_{\zeta_0}+\zeta_1\partial_{\zeta_1})+z(\zeta_0\partial_{\zeta_0}-\zeta_1\partial_{\zeta_1})+{2\over q}t\cr
    =&(-t^2+z^2)\partial_t+2t D+z(\zeta_0\partial_{\zeta_0}-\zeta_1\partial_{\zeta_1})\\
    \dD=&t\partial_t +z\partial_z +{1\over 2}\zeta_0\partial_{\zeta_0}+{1\over 2} \zeta_1\partial_{\zeta_1}+{1\over q} \\
    \dQ=&-{1\over 2} \zeta_0(\partial_t+\partial_z)+{1\over 2} \zeta_1(-\partial_t+\partial_z)+\partial_{\zeta_0}+\partial_{\zeta_1}\\
    \dS=&(t+z)\partial_{\zeta_0}-(-t+z)\partial_{\zeta_1}-{1\over 2}\zeta_0(t+z)(\partial_t+\partial_z)-{1\over 2}\zeta_1(-t+z)(-\partial_t+\partial_z)\cr
    &-{1\over q}(\zeta_0+\zeta_1)
\end{align}
and, the Casimir can be written as
\begin{align}
    &\mathcal{C}={1\over q^2}-{1\over 2q} +{2\over q} z\partial_z  +z^2(-\partial_t^2+\partial_z^2)-z\partial_t(\zeta_0 \partial_{\zeta_0}- \zeta_1\partial_{\zeta_1})+(z\partial_z+{1\over 2q})(\zeta_0\partial_{\zeta_0}+\zeta_1\partial_{\zeta_1}) \cr
    &+{1\over 2} \zeta_0\zeta_1 \partial_{\zeta_1}\partial_{\zeta_0} -z\partial_{\zeta_1}\partial_{\zeta_0}-{1\over 2q}\partial_z\zeta_0 \zeta_1  -{1\over 4z} (-z^2\partial_t^2+z^2\partial_z^2) \zeta_0\zeta_1 \cr
    &-({1\over 2}z\partial_z+{1\over 2q})(\zeta_0\partial_{\zeta_1}+\zeta_1\partial_{\zeta_0}) -{1\over 2} z\partial_t (\zeta_0\partial_{\zeta_1}- \zeta_1 \partial_{\zeta_0})
\end{align}
In the same way as in Section~\ref{sec: eigenfunction of casimir}, we obtain the following eigenfunctions of the Casimir:
\begin{itemize}
	\item {\bf Eigenvalue of Casimir:} $\boldsymbol{\nu\left(\nu-{1\over 2}\right)}$
	
	\begin{align}
      \Upsilon^1_{\nu w}=&e^{-iwt} z^{{1\over 2}-{1\over q} } J_\nu (wz) 
      \begin{pmatrix} 
      0 & -{\nu-\left({1\over 2}-{1\over q} \right)\over 2 z }  \\
      1   &0 \\
      \end{pmatrix} \\
       \Upsilon^2_{\nu w}=&e^{-iwt} z^{{1\over 2}-{1\over q} }J_{-\nu}(wz) \begin{pmatrix}
    0 &  -{\nu-\left({1\over 2}-{1\over q} \right)\over 2 z}  \\
    1 & 0 \\
    \end{pmatrix}\cr
    &\mbox{or}\quad \Upsilon^2_{\nu w}=e^{-iwt} z^{{1\over 2}-{1\over q} } Y_\nu (wz)\begin{pmatrix}
        0 & -{ \nu-\left({1\over 2}-{1\over q} \right)\over 2z} \\
        1 & 0\\
        \end{pmatrix} 
        \end{align}

	\item {\bf Eigenvalue of Casimir:} $\boldsymbol{\nu\left(\nu+{1\over 2}\right)}$
	
	\begin{align}
      \Upsilon^3_{\nu w}=&e^{-iwt} z^{{1\over 2}-{1\over q}} J_\nu (wz) 
      \begin{pmatrix} 
      0 & {\nu+\left({1\over 2}-{1\over q} \right)\over 2 z }  \\
      1   &0 \\
      \end{pmatrix} \\
       \Upsilon^4_{\nu w}=&e^{-iwt} z^{{1\over 2}-{1\over q}}J_{-\nu}(wz) \begin{pmatrix}
    0 &  {\nu+\left({1\over 2}-{1\over q} \right)\over 2 z}  \\
    1 & 0 \\
    \end{pmatrix}\cr
    &\mbox{or}\quad \Gamma^2_{\nu w}=e^{-iwt} z^{{1\over 2}-{1\over q}} Y_\nu (wz)\begin{pmatrix}
        0 & {\nu+\left({1\over 2}-{1\over q} \right) \over 2z} \\
        1 & 0\\
        \end{pmatrix}
    \end{align}

\end{itemize}
The other eigenfunctions are also similar to those in Appendix~\ref{app: casimir eigenfunction}. 

For general $q$ case, one can the collective action for the $\mathcal{N}=1$ SUSY SYK model is given by
\begin{equation}
S_{col}={N\over 2} \str \left[ - \smD\mstar  \Psi+ \log \Psi -{J\over q} \Psi  \mstar [\Psi]^{q-1} \right]\label{eq: bi-local collective action for SYK model general q}
\end{equation}
Note that the additional factor comes from the $i$'s in the action with disorder interaction which makes the Largrangian real. The large $N$ saddle point equation is given by
\begin{equation}
\mathbb{I} -J [\Psi]^{q-1}\mstar \Psi=0
\end{equation}
where we take the strong coupling limit. Using \eqref{eq: integral classical solution1} and \ref{eq: integral classical solution1}, one can easily evaluate the classical solution~\cite{Fu:2016vas} 
\begin{align}
\Psi_{cl}=c\begin{pmatrix}
0 & -{1\over q}  f^s_{1/q+1}(\tau_{12}) \\
f^a_{1/q}(\tau_{12}) & 0  \\
\end{pmatrix}\hspace{1cm} c=\left[{\tan {\pi \over 2q}\over 2\pi J}\right]^{1\over q}
\end{align}
and the eigenfunction of the quadratic action are found to be
\begin{align}
    u^1_{\nu w}(t,z)=&{1\over \sqrt{8\pi}}e^{-iwt} |J z|^{{1\over 2}-{1\over q}}Z_\nu^-(|wz|)\begin{pmatrix}
    0 & -{\nu -\left({1\over 2}-{1\over q}\right)\over 2|z|} \\
    \sgn(z) & 0 \\
    \end{pmatrix}\\
    u^2_{\nu w}(t,z)=&{1\over \sqrt{8\pi}}e^{-iwt} |J z|^{{1\over 2}-{1\over q}}Z_\nu^-(|wz|)\begin{pmatrix}
    0 & {\nu + \left({1\over 2}-{1\over q}\right)\over 2|z|} \\
    \sgn(z) & 0 \\
    \end{pmatrix}
\end{align}
We also confirm that 
\begin{align}
    \tilde{u}^1_{\nu w}(\tau_1,\tau_2)=&-{A_q\over \sqrt{8\pi}} J {e^{-{iw\over 2} (\tau_1+\tau_2)}\over \left|{J\over 2}(\tau_1-\tau_2)\right|^{{1\over 2}-{1\over q}}} Z^-_\nu(|{w\over 2}(\tau_1-\tau_2)|)\begin{pmatrix}
    0 & -{\nu +\left({1\over 2}-{1\over q}\right) \over 2\left|{1\over 2}(\tau_1-\tau_2)\right|}\sgn(\tau_1-\tau_2) \\
    1 & 0 \\
    \end{pmatrix}\cr
    \sim &\left[(\Psi_{cl})^{q-2} u_{\nu w}\right]
\end{align}
for some constant $A_q$.

\bibliographystyle{JHEP}
\bibliography{susysyk06}

\end{document}